\documentclass{article}

\usepackage{amsmath,amssymb,graphicx,amstext}
\usepackage{tikz}
\usepackage{caption}
\usepackage{subcaption}
\usepackage{textcomp}
\usepackage{booktabs}
\usepackage{changes}
\usepackage{rotating}
\usepackage{lineno,hyperref}
\usepackage{listings}
\usepackage{pdflscape}
\usepackage{amsthm}
\usepackage[top=2 cm,bottom=2 cm,left=2.5 cm, right=2.5 cm]{geometry}
\begin{document}

\title{Mathematical Modelling of Auxin Transport in Plant Tissues: Flux meets Signalling and Growth}

\author{Henry R. Allen\footnote{Department of Mathematics, Fulton Building, University of Dundee, Dundee, United Kingdom, DD1 4HN. Supported by EPSRC DTA PhD studentship}, Mariya Ptashnyk\footnote{Department of Mathematics, Colin Maclaurin Building, Heriot-Watt University, Edinburgh, United Kingdom, EH14 4AS, m.ptashnyk@hw.ac.uk}}

%\author{Henry R. Allen \and Mariya Ptashnyk}
%
%\institute{Henry R. Allen \at
%	       Department of Mathematics, Fulton Building, University of Dundee, Dundee, United Kingdom, DD1 4HN
%	     \and
%	     Mariya Ptashnyk \at
%	     Department of Mathematics, Colin Maclaurin Building, Heriot-Watt University, Edinburgh, United Kingdom, EH14 4AS, \\
%	   %  Tel.:  +44 (0)131 451 3214 \\
%	     \email{m.ptashnyk@hw.ac.uk}}

\maketitle

\begin{abstract}
Plant hormone auxin has critical roles in plant growth, dependent on its heterogeneous distribution in plant tissues. Exactly how auxin  transport and developmental processes such as growth coordinate to achieve the precise patterns of auxin observed experimentally is not well understood. Here we use mathematical modelling to examine the interplay between auxin dynamics and growth and their contribution to formation of patterns in auxin distribution in plant tissues. Mathematical models describing the auxin-related signalling pathway,  PIN and AUX1 dynamics, auxin transport, and cell growth in plant tissues are derived.     A key assumption of our models is the regulation of PIN proteins by the auxin-responsive ARF-Aux/IAA signalling pathway, with upregulation of PIN biosynthesis by ARFs.   Models are analysed and solved numerically to examine the long-time behaviour and auxin distribution. Changes in auxin-related signalling processes are shown to be able to trigger transition between passage and spot type patterns in auxin distribution. The model was also shown to be able to generate isolated cells with oscillatory dynamics in levels of  components of the auxin signalling pathway which could explain oscillations in levels of ARF targets that have been observed experimentally. Cell growth was shown to have influence on PIN polarisation and determination of auxin distribution patterns. Numerical simulation results indicate that auxin-related signalling processes can explain the different patterns in auxin distributions observed in plant tissues, whereas  the  interplay between auxin transport and growth can explain the `reverse-fountain' pattern in auxin distribution observed at plant root tips.

%\keywords{Mathematical modelling of signalling processes \and   Transport of hormone auxin  in plant tissues  \and  Plant growth and polarity of auxin-efflux carrier protein PIN}
\end{abstract}

%\linenumbers

%==============================================================================================
\section{Introduction}
%==============================================================================================

Plant growth and development is tightly controlled by the spatial distribution of the plant hormone auxin. Auxin distribution patterns are organ-specific \cite{Petersson_S_2009}, and may be classified into two general types: spot and passage patterns. Spot patterns are characterised by local maxima of auxin concentrations, and are observed in primordium initiation of leaves and flowers, as well as formation of lateral roots \cite{Benkova_E_2003,Dubrovsky_J_2008}. High concentration of auxin at these points promotes cell growth and division, leading to organ development. Passage patterns are characterised by files or networks of (neighbouring) cells which have higher auxin concentrations than those surrounding them, and are observed principally in the leaves and roots. In developing leaves, auxin distribution becomes arranged in a passage pattern forming networks, and the leaf veins are formed along these networks \cite{Biedron_M_2018}.

Auxin transport and distribution in a plant tissue are controlled by the family of auxin-efflux carrier protein PIN-FORMED (PIN) \cite{Leyser_O_2005} and auxin-influx carrier AUXIN RESISTANT1 (AUX1) \cite{Yang_Y_2006}. PIN is necessary for the formation of heterogeneous auxin distributions observed in plants \cite{Okada_K_1991}. PIN proteins are localised to the plasma membrane of cells where they are then responsible for active transport of auxin out of the symplast. While it is clear that some form of feedback mechanism exists that links auxin to the polarisation of PIN \cite{Chen_X_2014,Chen_X_2012,Robert_S_2010}, the exact nature of this feedback remains unclear \cite{Feng_M_2015,Gao_Y_2015}. One key hypothesis for the mode of the feedback mechanism is chemically via a so-called canalisation effect, where auxin flux through a cell membrane has a positive effect on PIN localisation to that membrane, however there is also evidence for a strain-based mechanism \cite{Homann_U_1998}, where PIN is localised to the membranes with higher mechanical strain. Differential expression of AUX1 is also required for auxin pattern formation in some tissues \cite{Swarup_R_2001}, however most cells  have symmetric distributions of AUX1 \cite{Kleine-Vehn_J-2008b}.

The dynamics and transport of auxin in a plant tissue are also regulated by an auxin-related cellular signalling pathway. Auxin influences gene expression via the so-called ARF-Aux/IAA signalling pathway \cite{Lavy_M_2016}. The signalling pathway describes a mechanism where auxin influences the levels of the family of gene transcription factors AUXIN RESPONSE FACTOR (ARF), via an interaction with Aux/IAA transcriptional repressors. Thus auxin modulates gene response by controlling the levels of ARFs, through which it plays a role in primary root growth \cite{Wan_J_2018}, root hair formation \cite{Zhang_DJ_2018}, fruit growth and flowering. It has also been shown that the auxin-related signalling pathway has an influence on PIN dynamics by having roles in governing its biosynthesis \cite{Paciorek_T_2005,Vieten_A_2005}, degradation \cite{Abas_L_2006}, and polarisation \cite{Sauer_M_2006}. Despite the clear importance of the ARF-Aux/IAA signalling pathway however, it is likely that this mechanism alone is not enough to explain all auxin responses and details of other auxin-related signalling processes  are emerging \cite{Leyser_O_2018}.  

Although the interactions between auxin, PIN and the auxin-related signalling pathway are essential for the transport and heterogeneous distribution of auxin in a plant tissue, which are necessary for growth and development of plants, the exact mechanism of nonlinear coupling between these processes is not yet completely understood. Thus the use of mathematical models to investigate the validity of possible interaction mechanisms is important to better understand the dynamics and pattern formation in auxin distribution in plant tissues. 

There are several results on mathematical modelling of auxin transport through plant tissues assuming that auxin influences the polarisation of PIN proteins in cell membranes. The flux-based transport enhancement approach (canalisation), where flux of auxin out of the cell through the membrane has a positive feedback on the localisation of PIN to this membrane, has been used to generate realistic branching patterns observed in leaf vein formation \cite{Feugier_F_2005,Fujita_2006}, and has also been analysed in \cite{Feller_C_2015,Stoma_S_2008}. When considering auxin transport through both apoplast and symplast, the auxin-dependent PIN distribution has been modelled by assuming that PIN proteins preferentially localise towards neighbouring cells with high auxin concentration. This approach was employed to generate spot-type patterns in auxin distribution observed in phyllotaxis \cite{Heisler_M_2006,Jonsson_H_2006} and auxin channels \cite{Merks_R_2007}. Further models considering influence of external auxin sensors on PIN distribution have also had success in capturing passage patterns in solutions of mathematical models including the apoplast \cite{Wabnik_K_2010}, although the biological relevance of this mechanism has been questioned \cite{Feng_M_2015,Gao_Y_2015}. The problem of generating different types of patterns in auxin distribution via unified mechanisms was addressed in \cite{Cieslak_M_2015} by considering the notion of `unidirectional fluxes' with a model based on petri nets and in \cite{Hayakawa_Y_2015} where the influence of non-flux-based feedback of auxin on PIN polarisation was described by auxin-dependent PIN degradation. Both of these models demonstrated that a change in a single parameter could lead to switching between passage and spot patterns in auxin distribution in a plant tissue. Mathematical models have also been used to show how the distribution of auxin in the plant root tip is maintained \cite{Band_L_2014,Mironova_V_2010}. An excellent summary of various mathematical models of polar auxin transport may be found in \cite{Berkel_K_2013}.

In this work we derive and analyse novel mathematical models for nonlinear interactions between auxin-related signalling processes, PIN and AUX1 dynamics, intercellular auxin flux, and growth of a plant tissue. 
For our modelling we primarily assume a flux-based mechanism of PIN localisation of a similar form as in \cite{Hayakawa_Y_2015}, coupled with a detailed model of the auxin signalling pathway.
We show that including the interplay between auxin-related signalling pathway and dynamics of PIN proteins in the mathematical model for auxin transport allows us to obtain both spot and passage type patterns in auxin distribution, depending on the values of the model parameter representing the rate of binding of PIN to auxin-TIR1.
Using linear stability analysis we determine the range of model parameters for which homogeneous patterns are stable.
This analysis identifies possible mechanisms for the formation of heterogeneous auxin distributions in plants and possible interaction points between auxin and PIN responsible for homogeneous, spot and passage patterns, respectively.
By considering model parameters that would generate oscillatory dynamics in auxin concentration in the model for auxin-related signalling pathway in a single cell, we show that the coupling between PIN dynamics, auxin transport,  and cellular signalling processes can explain the formation of oscillatory auxin responsiveness observed in the basal meristem of plant roots \cite{DeSmet_I_2007}.
Numerical simulations of the mathematical model for auxin transport, coupled with PIN dynamics, signalling processes and auxin-dependent growth,  suggest that cell growth can be one of the mechanisms underlying the formation of the `reverse fountain' of auxin flow in plant root tip \cite{Grieneisen_V_2007}.  Modelling and simulations of interactions between auxin-related signalling pathway and apoplastic auxin transport demonstrate dependence of pattern formation on assumptions on the mechanisms of auxin flux between symplast and apoplast and PIN localisation to the cell membrane. The incorporation of auxin-related signalling processes,  tissue growth and strain-dependent PIN polarisation into mathematical models for auxin transport,  analysis of oscillatory dynamics in auxin and PIN levels in  plant tissues and of formation of `reverse fountain' in  growing tissues,  as well as comparison between different mechanisms for auxin flux and PIN localization,  are the main novel contributions of the modelling and analysis presented here.

%revealed that in some cases  cellular signalling  process alone can account for formation of spot patterns in the  auxin distribution in plant tissues.

%\vspace{-0.2 cm } 

%==============================================================================================
\section{Materials and Methods}
%==============================================================================================

It is observed experimentally that cellular auxin mediates the dynamics of PIN  via its signalling pathway, 
whereas PIN regulates the heterogeneous distribution of auxin in tissues by controlling auxin flux between cells  \cite{Abas_L_2006,Paciorek_T_2005,Sauer_M_2006,Vieten_A_2005}. It is further known that auxin influences the plant growth on the cellular and organ level~\cite{Fendrych_M_2018,Reinhardt_D_2000}. 

In this work, we  derive and analyse new mathematical models for nonlinear interactions   between  auxin flux,  auxin-related signalling pathway, PIN and AUX1 dynamics,  and plant cell growth.   Incorporating the signalling and growth  processes  into mathematical models for auxin transport  allows us to investigate the influence of cellular processes on the distribution of auxin in plant tissues. 

%\vspace{-0.4 cm } 

%----------------------------------------------------------------------------------------------
\subsection*{Geometric Setting}
%----------------------------------------------------------------------------------------------

\begin{figure}[!ht]\centering
 \begin{tikzpicture}[scale=0.6]
  \draw[black,-] (0,5)--(0,0)--(5,0)--(10,0)--(10,5)--(5,5)--node[left]{$S^{m}_{3,4}$}node[right]{$S^{m}_{4,3}$}(5,0);
  \draw[black,-](5,5)--node[below]{$S^{m}_{3,1}$}node[above]{$S^{m}_{1,3}$}(0,5)--(0,10)--(5,10)--node[left]{$S^{m}_{1,2}$}node[right]{$S^{m}_{2,1}$}(5,5)--node[below]{$S^{m}_{4,2}$}node[above]{$S^{m}_{2,4}$}(10,5)--(10,10)--(5,10);
  \node (1) at (2.5,7.5){$V_{1}$};
  \node (2) at (7.5,7.5){$V_{2}$};
  \node (3) at (2.5,2.5){$V_{3}$};
  \node (4) at (7.5,2.5){$V_{4}$};
  
  \node (a) at (-0.5,10.5){\LARGE \textbf{a)}};
  
  \draw[black,-] (12,5)--(12,0)--(17,0)--(22,0)--(22,5)--(22,10)--(17,10)--(12,10)--(12,5)--(12.8,4.2)--(12.8,0.8)--(12,0);
  \draw[black,-](12.8,0.8)--(16.2,0.8)--(17,0)--(17,5)--(16.2,4.2)--node[left]{$S^{m}_{4,3}$}(16.2,0.8);
  \draw[black,-](16.2,4.2)--node[below]{$S^{m}_{3,1}$}(12.8,4.2)--(12,5)--(17,5)--(16.2,5.8)--node[above]{$S^{m}_{1,3}$}(12.8,5.8)--(12,5);
  \draw[black,-](12.8,5.8)--(12.8,9.2)--(12,10);
  \draw[black,-](12.8,9.2)--(16.2,9.2)--(17,10);
  \draw[black,-](16.2,9.2)--node[left]{$S^{m}_{1,2}$}(16.2,5.8)--(17,5)--(17,10)--(17.8,9.2)--(21.2,9.2)--(22,10);
  \draw[black,-](21.2,9.2)--(21.2,5.8)--(22,5)--(17,5)--(17.8,5.8)--node[right]{$S^{m}_{2,1}$}(17.8,9.2);
  \draw[black,-](17.8,5.8)--node[above]{$S^{m}_{2,4}$}(21.2,5.8)--(22,5)--(21.2,4.2)--(21.2,0.8)--(22,0);
  \draw[black,-](21.2,0.8)--(17.8,0.8)--(17,0);
  \draw[black,-](17.8,0.8)--node[right]{$S^{m}_{4,3}$}(17.8,4.2)--(17,5);
  \draw[black,-](17.8,4.2)--node[below]{$S^{m}_{4,2}$}(21.2,4.2);
  
  \node (c1) at (14.0,8.0){$V_{1}$};
  \node (c2) at (20.0,8.0){$V_{2}$};
  \node (c3) at (14.0,2.0){$V_{3}$};
  \node (c4) at (20.0,2.0){$V_{4}$};
  \node[rotate=-90] (a12) at (16.6,7.5){$V_{1,2}$};
  \node[rotate=90] (a21) at (17.45,7.5){$V_{2,1}$};
  \node (a13) at (14.5,5.4){$V_{1,3}$};
  \node (a24) at (19.5,5.4){$V_{2,4}$};
  \node (a31) at (14.5,4.6){$V_{3,1}$};
  \node (a42) at (19.5,4.6){$V_{4,2}$};
  \node[rotate=-90] (a34) at (16.6,2.5){$V_{3,4}$};
  \node[rotate=90] (a43) at (17.4,2.5){$V_{4,3}$};
  
  \node[rotate=-90] (s123) at (16.6,6){$S_{1}^{2,3}$};
  \node (s132) at (15.9,5.4){$S_{1}^{3,2}$};
  \node[rotate=90] (s214) at (17.4,6){$S_{2}^{1,4}$};
  \node (s241) at (18,5.4){$S_{2}^{4,1}$};
  \node (s314) at (16.2,4.6){$S_{3}^{1,4}$};
  \node[rotate=-90] (s341) at (16.6,4){$S_{3}^{4,1}$};
  \node (s423) at (18.1,4.6){$S_{4}^{2,3}$};
  \node[rotate=90] at (17.4,4.1){$S_{4}^{3,2}$};
   
  \node (b) at (11.5,10.5){\LARGE \textbf{b)}};
 \end{tikzpicture}
 \caption{Schematics of the tissue geometry used for numerical simulations. \textbf{a)} Simple geometry considering only intracellular space and cell membrane, with auxin flux considered to occur directly between cells.  Here $V_{i}$ represents the volume of cell $i$, and $S_{ij}^{m}$ represents the size of the portion of the membrane of cell $i$ between cells $i$ and $j$.  \textbf{b)} Schematics of a plant tissue where the domains representing the apoplast are equally divided between neighbouring cells and passive auxin flux also occurs in the apoplast.   Here $V_{i,j}$ represents the volume of apoplast compartment bordering cell $i$ between cells $i$ and $j$, and $S_{i}^{jk}$ represents the size of the border between apoplast compartments $(i,j)$ and $(i,k)$.}
 \label{fig:Geometry}
\end{figure}
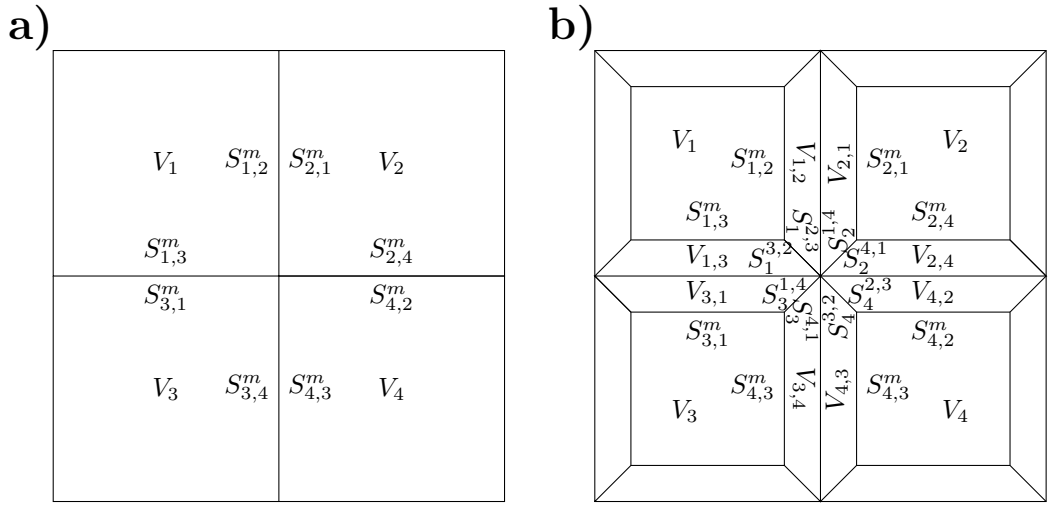

In our models for auxin dynamics a plant tissue is represented by a regular lattice of $N$ cells of square shape, and equal size and dimensions, Fig.~\ref{fig:Geometry}. In modelling auxin transport through a plant tissue we shall consider two cases: i) assuming direct interactions between neighbouring cells Fig.~\ref{fig:Geometry}~a) and ii) distinguishing between auxin dynamics in symplast and apoplast. In the second case we split the apoplast (middle lamella and plant cell walls) so that each cell has an equal portion of apoplast surrounding it. Then on a regular lattice the geometry of a plant tissue will be given by squares representing the cell inside, surrounded by four equal, regular trapeziums representing the apoplast, Fig.~\ref{fig:Geometry}~b). Similar geometric representations have been used in previous models \cite{Wabnik_K_2010}.

%\vspace{-0.4 cm } 
%----------------------------------------------------------------------------------------------
\subsection*{Mathematical Model for Auxin-related Signalling Pathway}
%----------------------------------------------------------------------------------------------

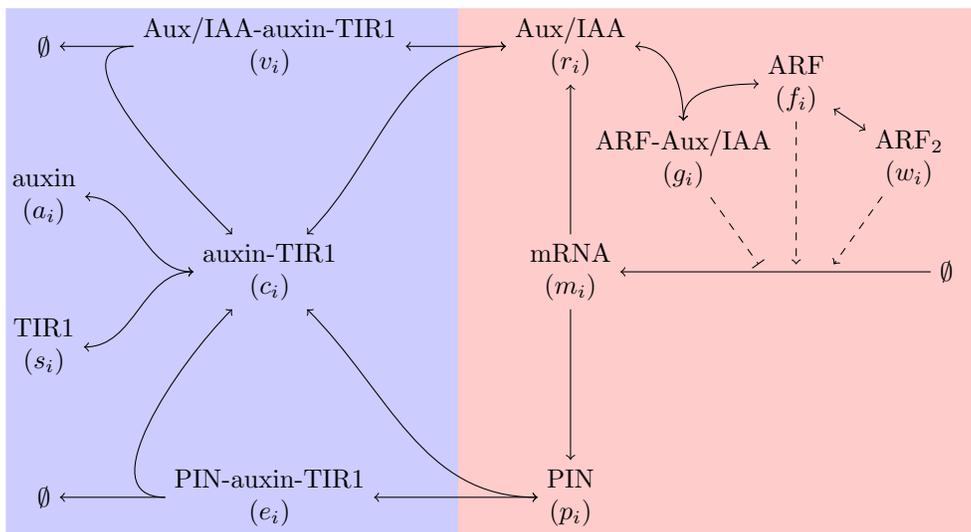
\begin{figure}[!ht]\centering
 \begin{tikzpicture}
  \filldraw[fill=red!20!white, draw=red!20!white] (4.5,3.5) -- (11.5,3.5) -- (11.5,-3.5) -- (4.5,-3.5) -- cycle;
  \filldraw[fill=blue!20!white, draw=blue!20!white] (-1.5,3.5) -- (4.5,3.5) -- (4.5,-3.5) -- (-1.5,-3.5) -- cycle;
  
  \node[align=center] (a) at (-1,1){auxin\\ ($a_{i}$)};
  \node[align=center] (s) at (-1,-1){TIR1\\ ($s_{i}$)};
  \node[align=center] (c) at (2,0){auxin-TIR1\\ ($c_{i}$)};
  \node[align=center] (r) at (6,3){Aux/IAA\\ ($r_{i}$)};
  \node[align=center] (p) at (6,-3){PIN\\ ($p_{i}$)};
  \node[align=center] (v) at (2,3){Aux/IAA-auxin-TIR1\\ ($v_{i}$)};
  \node[align=center] (e) at (2,-3){PIN-auxin-TIR1\\ ($e_{i}$)};
  \node[align=center] (emp1) at (-1,3){$\emptyset$};
  \node[align=center] (emp2) at (-1,-3){$\emptyset$};
  \node[align=center] (m) at (6,0){mRNA\\ ($m_{i}$)};
  \node[align=center] (f) at (9,2.5){ARF\\ ($f_{i}$)};
  \node[align=center] (w) at (10.5,1.5){ARF$_{2}$\\ ($w_{i}$)};
  \node[align=center] (g) at (7.5,1.5){ARF-Aux/IAA\\ ($g_{i}$)};
  \node[align=center] (emp3) at (11,0){$\emptyset$};
  
  \draw[black, <->] (a) to [out=0, in=180] (c);
  \draw[black, <->] (s) to [out=0, in=180] (c);
  \draw[black, <->] (r) -- (v);
  \draw[black, <->] (p) -- (e);
  \draw[black, <->] (r) to [out=180, in=45] (c);
  \draw[black, <->] (p) to [out=180, in=315]  (c);
  \draw[black, ->] (v) -- (emp1);
  \draw[black, ->] (e) -- (emp2);
  \draw[black, ->] (v) to [out=180, in=135] (c);
  \draw[black, ->] (e) to [out=180, in=225] (c);
  \draw[black, ->] (m) -- (r);
  \draw[black, ->] (m) -- (p);
  \draw[black, ->] (emp3) -- (m);
  \draw[black, <->] (r) to [out=0, in=90] (g);
  \draw[black, <->] (f) to [out=180, in=90] (g);
  \draw[black, <->] (f) -- (w);
  \draw[black, dashed, ->] (f) -- (9,0.1);
  \draw[black, dashed, -|] (g) -- (8.5,0.1);
  \draw[black, dashed, ->] (w) -- (9.5, 0.1);
 \end{tikzpicture}
 \caption{Schematic of the auxin-related signalling pathway. We assume PIN interacts with the auxin signalling pathway similar to Aux/IAA-type auxin-response proteins, hence PIN is degraded due to activation of the auxin-related signalling pathway.}
 \label{fig:signalling}
\end{figure}

In plant cells auxin is perceived by the TRANSPORT INHIBITOR RESPONSE 1 (TIR1) receptor protein \cite{Dharmasiri_N_2005,Kepinski_S_2005}. Upon perception by and direct binding to TIR1, auxin enhances the interactions between TIR1 and Aux/IAA by acting as a `molecular glue' \cite{Tan_X_2007}, and the enhanced interaction between TIR1s and Aux/IAAs leads to the degradation of Aux/IAAs \cite{Salehin_M_2015,Wang_R_2014}, see Fig.~\ref{fig:signalling}. When auxin concentrations are low, Aux/IAAs repress activity of AUXIN RESPONSE FACTOR (ARF)  by directly binding to ARFs and inhibiting their transcriptional ability. Correspondingly, a rise in auxin levels leads to degradation of Aux/IAAs, releasing  repression of ARFs by Aux/IAA. ARFs enhance the transcription of auxin-responsive mRNAs, including Aux/IAA, and may interact with the binding site as single monomers, dimers, and two monomers simultaneously. ARF binding sites may also interact with ARF-Aux/IAA complexes and   with  ARF and Aux/IAA as single molecules. Hence in the mathematical model, in each cell of a plant tissue we consider production and degradation of auxin, its binding to TIR1, and dissociation from TIR1. We also consider production of Aux/IAA from mRNA, Aux/IAA binding to and dissociation from auxin-TIR1, degradation of Aux/IAA from the Aux/IAA-auxin-TIR1 complex, and binding to and dissociation from ARF. We further consider dimerisation of ARF monomers and splitting of ARF dimers into ARF monomers. Finally we consider mRNA  transcription to be enhanced by ARFs as monomers, dimers, and double monomers, and inhibited by Aux/IAA as a single molecule interfering with an ARF monomer, and as the ARF-Aux/IAA complex. The total concentrations of TIR1 and ARF are considered to remain constant.

In \cite{Abas_L_2006} it was shown that auxin influences the degradation of PIN proteins via a mechanism similar to that by which auxin influences the degradation of Aux/IAAs. Auxin has also been shown to enhance PIN biosynthesis by controlling its gene expression through the ARF-Aux/IAA pathway \cite{Paciorek_T_2005,Vieten_A_2005}.
 From this limited evidence we will assume in our model that auxin feedback on PIN biosynthesis operates via the ARF-Aux/IAA signalling pathway, specifically that ARF upregulates mRNA encoding PIN.
We will assume that the mRNAs of Aux/IAA and PIN are identically regulated, and that PIN binds to the auxin-TIR1 complex whereupon it may be marked for degradation.
Thus low levels of auxin lead to auxin transport being inhibited due to repression of PIN biosynthesis, medium levels of auxin lead to increase in auxin transport due to enhanced biosynthesis of PINs, and high levels of auxin lead to its transport being inhibited due to enhanced degradation of PIN proteins, Fig.~\ref{fig:signalling}. Hence in the mathematical model, in each cell of a plant tissue we consider production of PIN from mRNA, association of PIN  to auxin-TIR1 and its dissociation  from auxin-TIR1, and PIN degradation from the PIN-auxin-TIR1 complex. 
  
Assuming spatial homogeneity of signalling processes in each cell, the dynamics of auxin signalling pathway on the cell level can be described by a system of ordinary differential equations 

\begin{equation}\label{eq:aux_flux-signal_cell}
 \begin{aligned}
  \frac{dm_{i}}{dt} = \; & \alpha_{m}\dfrac{\phi_{m}f_{i}/\theta_{f} + w_{i}/\theta_{w} + f_{i}^{2}/\psi_{f}}{1 + f_{i}/\theta_{f} + w_{i}/\theta_{w} + g_{i}/\theta_{g} + f_{i}r_{i}/\psi_{g} + f_{i}^{2}/\psi_{f}} - \mu_{m}m_{i},
  \\
  \frac{dr_{i}}{dt} = \; & \alpha_{r}m_{i} - \beta_{r}r_{i}c_{i} + \gamma_{r}v_{i} - \beta_{g}r_{i}f_{i} + \gamma_{g}g_{i},
  \\
  \frac{ds_{i}}{dt} = \; & -\beta_{a}a_{i}s_{i} + \gamma_{a}c_{i},
  \\
  \frac{dc_{i}}{dt} = \; & \beta_{a}a_{i}s_{i} - \gamma_{a}c_{i} + \left(\gamma_{r} + \mu_{r}\right)v_{i} - \beta_{r}r_{i}c_{i} + \left(\gamma_{p}+\mu_{p}\right)e_{i} - \beta_{p}p_{i}c_{i},
  \\
  \frac{dv_{i}}{dt} = \; & \beta_{r}r_{i}c_{i} - \left(\gamma_{r} + \mu_{r}\right)v_{i},
  \\
  \frac{de_{i}}{dt} = \; & \beta_{p}p_{i}c_{i} - \left(\gamma_{p}+\mu_{p}\right)e_{i},
  \\
  \frac{df_{i}}{dt} = \; & -2\beta_{f}f_{i}^{2} + 2\gamma_{f}w_{i} - \beta_{g}r_{i}f_{i} + \gamma_{g}g_{i},
  \\
  \frac{dg_{i}}{dt} = \; & \beta_{g}r_{i}f_{i} - \gamma_{g}g_{i},
  \\
  \frac{dw_{i}}{dt} = \; & \beta_{f}f_{i}^{2} - \gamma_{f}w_{i},
 \end{aligned}
\end{equation}
completed with  initial conditions given by initial concentrations of signalling molecules, specified in the analysis and numerical simulations of the mathematical model below, where the subscript $i$ denotes to which cell the variable belongs, $1 \leq i \leq N$, and  $N$ is the total number of cells. Here mRNAs are denoted by $m_{i}$, cytosolic PIN is denoted by $p_{i}$, auxin is denoted by $a_{i}$, Aux/IAA is denoted by $r_{i}$, TIR1 is denoted by $s_{i}$, auxin-TIR1 complex is denoted by $c_{i}$, PIN-auxin-TIR1 is denoted by $e_{i}$, Aux/IAA-auxin-TIR1 is denoted by $v_{i}$, ARF monomers are denoted by $f_{i}$, ARF-Aux/IAA complexes are denoted by $g_{i}$, and ARF\textsubscript{2} dimers are denoted by $w_{i}$.  A list of all variables considered in our models can be found in Table~\ref{tab:variables}.  Model \eqref{eq:aux_flux-signal_cell} is similar to the model for auxin signalling pathway derived in \cite{Middleton_A_2010}, with the inclusion of PIN as a secondary auxin response protein.

Parameter $\alpha_{m}$  is the rate of mRNA production, $\mu_{m}$ is the rate of  mRNA degradation, $\phi_{m}$ is the ratio of ARF-dependent mRNA production to ARF\textsubscript{2}- and double ARF-dependent mRNA production, and $\theta_{f}$, $\theta_{w}$, $\theta_{g}$, $\psi_{g}$, and $\psi_{f}$ are the binding thresholds to the relevant binding site of ARF monomers, ARF dimers, ARF--Aux/IAA complexes,  molecules of ARF and  Aux/IAA, and two molecules of ARF.  The rate of Aux/IAA translation is $\alpha_{r}$, whereas $\beta_{r}$ and $\gamma_{r}$ are the binding and dissociation rates of Aux/IAA and auxin-TIR1, $\beta_{g}$ and $\gamma_{g}$ are the binding and dissociation  rates of Aux/IAA and ARF, and $\mu_{r}$ is the degradation rate of Aux/IAA from Aux/IAA-auxin-TIR1.  By $\beta_{a}$ and $\gamma_{a}$  the binding and dissociation rates of auxin and TIR1 are denoted,  whereas  $\beta_{f}$ and $\gamma_{f}$ are the binding and dissociation rates of two ARF proteins,   $\beta_{p}$ and $\gamma_{p}$ are the binding and dissociation rates of PIN and auxin-TIR1, and $\mu_{p}$ is the rate of degradation of PIN from the PIN-auxin-TIR1 complex. 

%----------------------------------------------------------------------------------------------
\subsection*{Auxin Transport in Plant Tissues}
%----------------------------------------------------------------------------------------------

In the mathematical model for auxin transport in a plant tissue, we consider the dynamics of cellular auxin $a_i$, the PIN-mediated flux of auxin between neighbouring cells, and the dynamics of cellular $p_i$ and membrane-bound  $P_{ij}$ PIN.  The index $ij$ denotes the membrane of cell $i$ between two neighbouring cells $i$ and $j$, e.g.\ $S^{m}_{ij}$ denotes the size of the portion of the membrane of cell $i$ between cells $i$ and $j$. 

Auxin $a_i$ is produced inside the cells with rate $\alpha_{a}$, degraded with rate $\mu_{a}$, and transported between cells by membrane-bound PIN $P_{ij}$. Cellular PIN $p_i$ is translated from mRNAs with rate $\alpha_{p}$ and its localisation to the cell membrane depends on the auxin flux through the membrane: stronger auxin flux through a specific membrane portion enhances localisation and leads to higher concentration of membrane-bound PIN $P_{ij}$ in that part of the cell membrane.

Considering homogeneous distribution of membrane-bound PIN on each part of a cell membrane, see Fig.~\ref{fig:Geometry}, the interplay between auxin flux and PIN dynamics is modelled by a system of strongly coupled nonlinear ODEs 

\begin{equation}\label{eq:aux_flux-signal_membrane}
 \begin{aligned}
  \frac{da_{i}}{dt} = \; & \alpha_{a} + \gamma_{a}c_{i} - \beta_{a}a_{i}s_{i} - \mu_{a}a_{i} - \dfrac{1}{V_{i}}\sum_{i\sim j}S^{m}_{ij}J_{a}^{ij},
  \\
  \frac{dp_{i}}{dt} = \; & \alpha_{p}m_{i} - \beta_{p}p_{i}c_{i} + \gamma_{p}e_{i} -\dfrac{1}{V_{i}}\sum_{i\sim j}S^{m}_{ij}J_{p}^{ij},
  \\
  \frac{dP_{ij}}{dt} = \; & J_{p}^{ij},
 \end{aligned}
\end{equation}
where $i\sim j$ is short notation for $j \in \{k \ | \ \text{cell} \ i \ \text{neighbours cell} \ k\}$ and $V_i$ denotes the volume of the cell $i$. The flux of auxin $J_{a}^{ij}$ between neighbouring cells $i$ and $j$ and the localisation of cytosolic PIN $p_i$ from cell $i$ to membrane portion $ij$ facing cell $j$ together with dissociation of membrane-bound PIN $P_{ij}$ back to the cell $J_{p}^{ij}$  are given by 

\begin{equation}\label{eq:aux_flux-signal_fluxes}
 \begin{aligned}
  J_{a}^{ij} = \; & \phi_{A}\left(a_{i}P_{ij} - a_{j}P_{ji}\right), 
  \\
  J_{p}^{ij} = \; &  \lambda p_{i}H\left(J_{a}^{ij}\right) - \delta_{p}P_{ij}, 
   \\
\text{where }  &  H(J_{a}^{ij}; \lambda) = \;  \dfrac{\frac{1}{1 + \exp\{-h\left(J_{a}^{ij}/\lambda - \theta\right)\} }}{\sum_{i\sim k}\frac{1}{1 + \exp\{-h\left(J_{a}^{ik}/{\lambda} - \theta\right)\} }}.
 \end{aligned}
\end{equation}

Here $H$ is a function describing the feedback of auxin flux on PIN localisation and is defined such that it is bounded between $0$ and $1$, increasing in $J_{a}^{ij}$, and $\sum_{i\sim j} H\left(J_{a}^{ij}; \lambda\right) = 1$. Parameter $\phi_{A}$ denotes the rate of PIN-mediated auxin transport, $\lambda$ is the maximal rate of PIN localisation to the membrane, $\delta_{p}$ denotes the rate of PIN dissociation from the membrane, $h$ is the flux-response coefficient, and $\theta$ is the flux threshold for positive feedback. In the response term $H$ flux is scaled by  $\lambda$. The individual response terms $1/\left[1 + \exp\{-h\left(J_{a}^{ij}/{\lambda} - \theta\right)\}\right]$ were chosen such that for fluxes smaller (greater) than a threshold value $\theta$, i.e.~$J_{a}^{ij}/{\lambda} < (>)\ \theta$, the individual response would be approximately zero (one), ensuring strong positive feedback for large auxin fluxes.

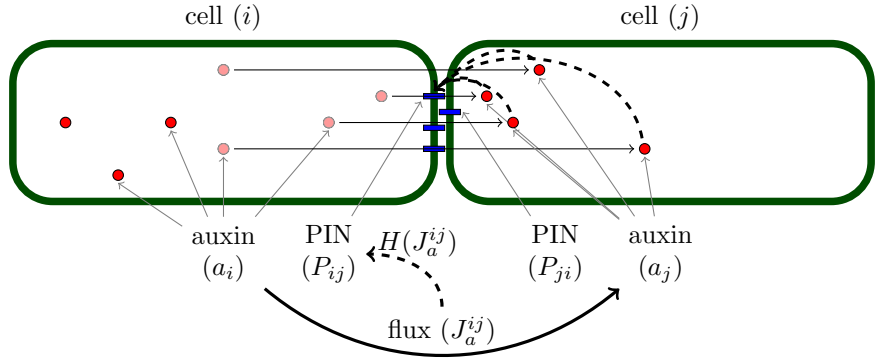
\begin{figure}[!ht]\centering
 \begin{tikzpicture}[scale=0.7]
  \node at (4,3.5){cell $(i)$};
  \node at (12.3,3.5){cell $(j)$};
  \filldraw[rounded corners=15pt, fill=white, draw=green!30!black, line width=1mm] (0,0) rectangle (8,3);
  \filldraw[rounded corners=15pt, fill=white, draw=green!30!black, line width=1mm] (8.3,0) rectangle (16.3,3);
   
  \filldraw[draw=black,fill=red] (4,1) circle (0.1);
  \filldraw[draw=black,fill=red] (2,0.5) circle (0.1);
  \filldraw[draw=black,fill=red] (6,1.5) circle (0.1);
  \filldraw[draw=black,fill=red] (3,1.5) circle (0.1);
  \filldraw[draw=black,fill=red] (7,2) circle (0.1);
  \filldraw[draw=black,fill=red] (1,1.5) circle (0.1);
  \filldraw[draw=black,fill=red] (4,2.5) circle (0.1);
   
  \node[align=center] (auxin) at (4,-1){auxin\\ ($a_{i}$)};
  \draw[gray,->] (auxin) -- (2,0.35);
  \draw[gray,->] (auxin) -- (6,1.35);
  \draw[gray,->] (auxin) -- (4,0.85);
  \draw[gray,->] (auxin) -- (3,1.35);
  
  \filldraw[draw=black,fill=blue] (7.8,1.95) rectangle (8.2,2.05);
  \node[align=center] (PIN) at (6,-1){PIN\\ ($P_{ij}$)};
  \draw[gray,->] (PIN) -- (7.75,1.9);
  \filldraw[draw=black,fill=blue] (8.1,1.65) rectangle (8.5,1.75);
  \node[align=center] (PIN2) at (10.3,-1){PIN\\ ($P_{ji}$)};
  \draw[gray,->] (PIN2) -- (8.55,1.7);
   
  \filldraw[draw=black,fill=red] (9,2) circle (0.1);
  \filldraw[draw=black!40!white,fill=red!40!white] (7,2) circle (0.1);
  \draw[black,->] (7.2,2) -- (8.8,2);
  \node[align=center] (auxin2) at (12.3,-1){auxin\\ ($a_{j}$)};
  \draw[black,->, very thick] (auxin) to[out=320,in=220]node[above]{flux ($J_{a}^{ij}$)} (auxin2);
   
  \draw[gray,->] (auxin2) -- (9,1.85);
  
  \draw[black,dashed,->,line width=0.4mm] (8.9,2.2) to[out=135,in=45] (8,2.1);
  \filldraw[draw=black,fill=blue] (7.8,0.95) rectangle (8.2,1.05);
  \filldraw[draw=black,fill=blue] (7.8,1.35) rectangle (8.2,1.45);
  \draw[black,dashed,->,very thick] (8.15,-2) to[out=90,in=0]node[above]{$H(J_{a}^{ij})$} (PIN);
   
  \filldraw[draw=black!40!white,fill=red!40!white] (4,1) circle (0.1);
  \filldraw[draw=black!40!white,fill=red!40!white] (6,1.5) circle (0.1);
  \filldraw[draw=black!40!white,fill=red!40!white] (4,2.5) circle (0.1);
  \filldraw[draw=black,fill=red] (12,1) circle (0.1);
  \filldraw[draw=black,fill=red] (9.5,1.5) circle (0.1);
  \filldraw[draw=black,fill=red] (10,2.5) circle (0.1);
  \draw[black,->] (4.2,1) -- (11.8,1);
  \draw[black,->] (6.2,1.5) -- (9.3,1.5);
  \draw[black,->] (4.2,2.5) -- (9.8,2.5);
  \draw[gray,->] (auxin2) -- (12,0.85);
  \draw[gray,->] (auxin2) -- (9.5,1.35);
  \draw[gray,->] (auxin2) -- (10,2.35);
  \draw[black,dashed,->,line width=0.4mm] (11.95,1.2) to[out=100,in=45] (8,2.1);
  \draw[black,dashed,->,line width=0.4mm] (9.45,1.7) to[out=110, in=45] (8,2.1);
  \draw[black,dashed,->,line width=0.4mm] (9.9,2.7) to[out=145,in=45] (8,2.1);
  
 \end{tikzpicture}
 \caption{Schematic of PIN-mediated auxin transport between two cells. Auxin (red circles) is transported from cell $i$ to cell $j$ by the efflux protein PIN (blue rectangles). In mathematical models, the concentration of auxin in cell $i$ is denoted by $a_{i}$, and the concentration of PIN localised to the portion of cell \textit{i}'s membrane which neighbours cell $j$ is denoted $P_{ij}$. The flux of auxin from cell $i$ to cell $j$ is denoted by $J_{a}^{ij}$, and is assumed to positively feedback on the localisation of PIN to membrane portion $ij$ between cells $i$ and $j$.}
 \label{fig:transport}
\end{figure}

%----------------------------------------------------------------------------------------------
\subsection*{Auxin-dependent Tissue Growth}
%----------------------------------------------------------------------------------------------

It has been observed that auxin can enhance cell growth in shoots \cite{Reinhardt_D_2000}, however auxin can inhibit primary root cell elongation \cite{Overvoorde_P_2010}. Thus in our model we consider the growth rate of a cell to be dependent on the concentration of auxin within the cell 
\begin{subequations}\label{eq:growth}
 \begin{align}
  \dfrac{dx_{i}}{dt} = & \; \chi\dfrac{a_{i}}{\theta_{x} + a_{i}}\prod_{i\sim j, i\| j}\dfrac{x_{j}}{x_{i}}, 
 \label{eq:growth_a} \\
  \dfrac{dx_{i}}{dt} = & \; \chi\dfrac{\theta_{x}}{\theta_{x} + a_{i}}\prod_{i\sim j, i\| j}\dfrac{x_{j}}{x_{i}}, \label{eq:growth_b}
 \end{align}
\end{subequations}
where $x_{i}$ denotes the length of either the horizontal or vertical wall of cell $i$, $\chi$ is the maximum growth rate, and $\theta_{x}$ is the threshold for half-maximal auxin-dependent growth rate. Here $i\| j$ denotes that if $x_{i}$ is the horizontal (vertical) length of cell $i$ then $x_{j}$ is the horizontal (vertical) length if cell $j$. The first equation \eqref{eq:growth_a} corresponds to auxin-enhanced growth, whereas equation \eqref{eq:growth_b} describes  auxin-inhibited growth.

The growth of a cell is constrained by the cell wall and adhesion between cells leading to `tissue tension', where slow growing neighbouring cells will constrain growth of the cell, and fast growing neighbouring cells will accelerate its growth. Hence in our model we include a simple term for tissue tension such that growth rate of a cell is scaled by the ratio of the neighbouring cell length to the current cell length. 

When considering signalling processes and auxin and PIN dynamics in a growing tissue, equations \eqref{eq:aux_flux-signal_cell}-\eqref{eq:aux_flux-signal_fluxes} are modified by including the dilution effect due to growth: 
\begin{equation*}
 \begin{aligned}
  \dfrac{dy_{i}}{dt} = & \; F_{i} - \frac{1}{V_{i}}\dfrac{dV_{i}}{dt}y_{i},
  \\
  \dfrac{dY_{ij}}{dt} = & \; F_{ij} - \frac{1}{S^{m}_{ij}}\dfrac{dS^{m}_{ij}}{dt}Y_{ij},
 \end{aligned}
\end{equation*}
where $y_{i}$ ($Y_{ij}$) denotes concentration of a chemical in cell $i$ (membrane $ij$), and $F_{i}$ ($F_{ij}$) denotes the reaction terms in the corresponding equations. Since in our model the cell shapes are simplified to be rectangular, $S^{m}_{ij}$ is taken to be the cell length along the appropriate axis, and $V_{i}$ is the product of the length and width of the cell, where $\dfrac{dV_{i}}{dt}$ and $\dfrac{dS^{m}_{ij}}{dt}$ are determined by \eqref{eq:growth} for the corresponding sides of the cell $i$.

%----------------------------------------------------------------------------------------------
\subsection*{Strain-dependent PIN Localisation}
%----------------------------------------------------------------------------------------------

There is evidence that plasma membranes undergoing higher strains have increased PIN localisation to them \cite{Homann_U_1998}. We model this mechanism by considering PIN localisation depending on the strain rate of the corresponding cell membrane

\begin{equation}\label{eq:strain}
 J_{p}^{ij} = \left(\lambda H\left(J_{a}^{ij}; \lambda+\nu\right) + \nu\frac{1}{x_{i}}\dfrac{dx_{i}}{dt}\right)p_{i} - \delta_{p}P_{ij}, 
\end{equation}
in addition to the auxin flux-related PIN localisation (compare with \eqref{eq:aux_flux-signal_fluxes}), where $\nu$ is the strain-dependent rate of PIN localisation to the cell membrane. Here in the response term $H$ flux is scaled by $\lambda+\nu$.

%----------------------------------------------------------------------------------------------
\subsection*{Symplast-Apoplast Model for Auxin Transport in Plant Tissues}
%----------------------------------------------------------------------------------------------

Mathematical models considering  direct flux of auxin between cells (see e.g.~\cite{Hayakawa_Y_2015,Stoma_S_2008})  provide a good framework to analyse the auxin transport through a plant tissue. However along with active transport of auxin in/out of the cell it is important to consider the effect of passive flux of auxin through the apoplast. As described above, auxin is transported out of the cell symplast by membrane-bound PIN proteins. Due to the pH gradient between the apoplast and cytoplasm and weakly acidic nature of auxin, auxin passively diffuses from the cell wall into cell interiors, however auxin is  transported into cell symplast by membrane-bound influx proteins AUX1 at a much higher rate \cite{Rubery_P_1974,Yang_Y_2006}. Auxin influx protein AUX1 is synthesised within cells and then is trafficked to the cell membrane. Biosynthesis of AUX1 is known to be enhanced by auxin \cite{Heisler_M_2006}. Contrasting PIN, AUX1 is symmetrically localised in membranes for most plant cells \cite{Kleine-Vehn_J-2008b}.

Thus when considering both symplast (cell inside) and apoplast (plant cell walls and middle lamella),  the mathematical model for auxin transport through a plant tissue, coupled with cellular signalling processes and dynamics of PIN and AUX1, in addition to equations \eqref{eq:aux_flux-signal_cell} and new equations for $a_i$, $p_i$, and $P_{ij}$,  includes the dynamics of auxin $A_{ij}$ in apoplast, cellular AUX1 $u_{i}$, and membrane-bound AUX1 $U_{ij}$:

\begin{equation}\label{eq:Ap-AUX-cell}
 \begin{aligned}
  \frac{da_{i}}{dt} = \; & \alpha_{a} + \gamma_{a}c_{i} - \beta_{a}a_{i}s_{i} - \mu_{a}a_{i} - \dfrac{1}{V_{i}}\sum_{i\sim j}S^{m}_{ij}J_{a}^{ij},
  \\
  \frac{dA_{ij}}{dt} = \; & \dfrac{1}{V_{ij}}\Big(S_{ij}^{m}J_{a}^{ij} - S_{ij}^{w}J_{A}^{ij} - \sum_{j\sim k}S_{i}^{jk}J_{i}^{jk}\Big) - \mu_{a}A_{ij},
  \\
  \frac{dp_{i}}{dt} = \; & \alpha_{p}m_{i} - \beta_{p}p_{i}c_{i} + \gamma_{p}e_{i} -\dfrac{1}{V_{i}}\sum_{i\sim j}S^{m}_{ij}J_{p}^{ij},
  \\
  \frac{dP_{ij}}{dt} = \; & J_{p}^{ij},
  \\
  \frac{du_{i}}{dt} = \; & \alpha_{u}m_{i} - \mu_{u}u_{i}- \frac{1}{V_{i}}\sum_{i\sim j}S_{ij}^{m}J_{u}^{ij},
  \\
  \frac{dU_{ij}}{dt} = \; & J_{u}^{ij}, 
 \end{aligned}
\end{equation}
where $J_{a}^{ij}$ is the flux of auxin between  cell $i$ and $ij$-part of the apoplast, and $J_{u}^{ij}$ denotes the localisation of AUX1 from cell $i$ to membrane portion $ij$ together with dissociation of $U_{ij}$ back to the cell.  Parameter $\alpha_{u}$ is the translation rate of AUX1 from mRNA, and $\mu_{u}$ is the degradation rate of AUX1. A list of all variables considered in our models can be found in Table~\ref{tab:variables}.

The transport of auxin across the $ij$-part of plasma membrane (part of membrane between cell $i$ and $ij$-part of apoplast) combines active transport by PIN and AUX1 and a small contribution from passive diffusion. In apoplast we consider passive diffusion of auxin between neighbouring apoplast compartments, denoted by $ J_{A}^{ij}$ and $ J_{i}^{jk}$ for different parts of apoplast,  and where $S_{ij}^{w}$ denotes the size of the interface between apoplast compartments $ij$ and $ji$, $S_{i}^{jk}$ denotes the size of the interface between apoplast compartments $ij$ and $ik$, and $V_{ij}$ denotes the size of apoplast compartment $ij$.  The passive fluxes of auxin through the apoplast and of AUX1 localisation to the membrane are given by

\begin{equation}\label{eq:Ap-pas}
 \begin{aligned} 
  J_{A}^{ij} = \; & \phi_{A}\left(A_{ij} - A_{ji}\right),
  \\
  J_{i}^{jk} = \; & \phi_{A}\left(A_{ij} - A_{ik}\right),
  \\
  J_{u}^{ij} = \; & \omega_{u}u_{i} - \delta_{u}U_{ij}. 
 \end{aligned}
\end{equation}
Here $\phi_{A}$ is the rate of passive flux of auxin through the apoplast, $\omega_{u}$ is the rate of AUX1-membrane localisation, and $\delta_{u}$ denotes the rate of AUX1-membrane dissociation. To analyse the emergence of patterns in auxin distribution  and PIN polarisation in plant tissues we shall compare two different types of auxin transport and PIN localisation:
\begin{subequations}\label{eq:Ap-Fluxes}
 \begin{align}
  J_{a}^{ij} = \; & \phi_{a}\left(\kappa_{a}^{ef}a_{i} - \kappa_{a}^{in}A_{ij}\right) + \phi_{p}P_{ij}\left(\kappa_{p}^{ef}\dfrac{a_{i}}{\theta_{a}^{p} + a_{i}} - \kappa_{p}^{in}\dfrac{A_{ij}}{\theta_{a}^{p} + A_{ij}}\right) \nonumber\\
  & \qquad \qquad \ \; + \; \phi_{u}U_{ij}\left(\kappa_{u}^{ef}\dfrac{a_{i}}{\theta_{a}^{u} + a_{i}} - \kappa_{u}^{in}\dfrac{A_{ij}}{\theta_{a}^{u} + A_{ij}}\right),
 \label{eq:Ap-Fluxes_a} \\
  \tilde{J}_{a}^{ij} = \; & \phi_{a}\left(\kappa_{a}^{ef}a_{i} - \kappa_{a}^{in}A_{ij}\right) + \tilde{\phi}_{p}P_{ij}\kappa_{p}^{ef}a_{i} - \tilde{\phi}_{u}U_{ij}\kappa_{u}^{in}A_{ij}, 
 \label{eq:Ap-Fluxes_b} \\
  J_{p}^{ij} = \; & \omega_{p}\left( (1-\kappa_{p}) + \kappa_{p}\frac{a_{j}}{\theta_{p}^{a} + a_{j}}\right)p_{i} - \delta_{p}P_{ij}, \label{eq:Ap-Fluxes_c}
  \\
  \tilde{J}_{p}^{ij} = \; & \omega_{p} H\left(J\right)p_{i} - \delta_{p}P_{ij}, \label{eq:Ap-Fluxes_d}
 \end{align}
\end{subequations}
where $J_{a}^{ij}$ features saturating auxin transport \cite{Heisler_M_2006,Jonsson_H_2006}, $\tilde J_{a}^{ij}$ is an extension of the flux considered in \eqref{eq:aux_flux-signal_fluxes} by including  the presence of apoplast, $J_{p}^{ij}$ is a mechanism for PIN localisation, proposed in e.g.~\cite{Heisler_M_2006,Jonsson_H_2006}, which, along with spontaneous localisation, specifies that higher auxin concentrations in neighbouring cells will cause PIN localisation to the membranes of the neighbouring cells, and $\tilde{J}_{p}^{ij}$ is the mechanism for PIN localisation  considered in \eqref{eq:aux_flux-signal_fluxes}, where $J$ denotes the mechanism for auxin flux given by  either \eqref{eq:Ap-Fluxes_a} or \eqref{eq:Ap-Fluxes_b}, depending on which flux is considered in numerical simulations of the model.

Here $\phi_{a}$ is the rate of passive flux of auxin through the cell membrane, $\phi_{p}$ is the rate of PIN-dependent saturating auxin flux, $\phi_{u}$ is the rate of AUX1-dependent saturating auxin flux, $\tilde{\phi}_{p}$ is the rate of PIN-dependent non-saturating auxin flux, and $\tilde{\phi}_{u}$ is the rate of AUX1-dependent non-saturating auxin flux. Parameters $\kappa_{a}^{ef}$, $\kappa_{p}^{ef}$, $\kappa_{u}^{ef}$ denote the passive, PIN-dependent, and AUX1-dependent efflux of auxin respectively, and $\kappa^{in}$, $\kappa_{p}^{in}$, $\kappa_{u}^{in}$ denote the passive, PIN-dependent, and AUX1-dependent influx of auxin respectively. Parameters $\theta_{a}^{p}$, $\theta_{a}^{u}$ denote the concentration of auxin for half-maximal transport by PIN and AUX1 respectively. In localisation processes, $\omega_{p}$ is the rate of PIN-membrane localisation, and $\delta_{p}$ is the rate of PIN-membrane dissociation.  Parameter $\kappa_p$ denotes the proportion of PIN localisation that is auxin-dependent and $\theta_p^a$ is the half-maximal concentration of auxin for auxin-dependent PIN localisation. In $J_{p}^{ij}$ the dependence of PIN localisation on concentrations of auxin in neighbouring cells may be related to the fact that auxin-enhanced  cell expansion places strain on the neighbouring membrane and thus enhances the PIN localisation \cite{Homann_U_1998}.% or to the auxin transport through plasmadesmata {\color{red} do we have a citation for this? }.

%----------------------------------------------------------------------------------------------
\subsection*{Numerical Methods and Implementation of Model Equations}
%----------------------------------------------------------------------------------------------

Numerical codes for  simulations of  model equations  \eqref{eq:aux_flux-signal_cell}-\eqref{eq:aux_flux-signal_fluxes} or   \eqref{eq:aux_flux-signal_cell},  \eqref{eq:Ap-AUX-cell}-\eqref{eq:Ap-Fluxes}  are implemented  in Python, taking advantage of the Scipy module \cite{SciPy}. Solutions were obtained using the scipy.integ\-ra\-te.odeint package which solves systems of ODEs using lsoda from the FORTRAN library odepack which can automatically select to use Adams (stiff) or BDF (non-stiff) methods, dynamically monitoring data to decide which method should be used \cite{Hindmarsh_A_1983,Petzold_L_1983}. 

For numerical simulations we consider two types of initial conditions: (i) small perturbations of the homogeneous steady state or (ii) zero concentrations for most molecules with the exception of TIR1 ($s_{i}$) and ARF ($f_{i}$) since the total amounts of TIR1 and ARF are conserved, and the conserved quantities were chosen as initial conditions. 
To calculate small perturbations around the homogeneous steady state, the homogeneous steady state was first calculated numerically and then in each cell the concentration of each component of the steady state solution was multiplied by a random number between $0.9$ and $1.1$.

For certain simulations we consider some cells to be either source or sinks. Compared to standard cells in the domain, in source cells the rate of auxin production $\alpha_{a}$ is doubled and in sink cells the rate of auxin degradation $\mu_{a}$ is doubled.
We solve symplast model \eqref{eq:aux_flux-signal_cell}-\eqref{eq:aux_flux-signal_fluxes} with  initial condition as a perturbation of homogeneous steady-state and periodic boundary condition to examine the emergence of spot and passage patterns in auxin distribution in plant tissue,  Fig.~\ref{fig:Basic_Sig2-Periodic-Spatial-7014}.
Zero-flux boundary conditions were considered to analyse the effect of boundary conditions on pattern formation. 
We solve model \eqref{eq:aux_flux-signal_cell}-\eqref{eq:growth_a} with  initial condition as a perturbation of the homogeneous steady-state and periodic boundary condition to examine the effect of tissue growth on the emergence of spot and passage patterns, Figs.~\ref{fig:Sig_horizontal},~\ref{fig:Sig_vertical}, and model \eqref{eq:aux_flux-signal_cell}-\eqref{eq:growth_a},\eqref{eq:strain} to examine how varying the weighting between flux-induced and strain-induced PIN localisation affects auxin distribution in spot and passage patterns, Fig.~\ref{fig:Sig_mech}.
 We solve symplast model \eqref{eq:aux_flux-signal_cell}-\eqref{eq:aux_flux-signal_fluxes} with  initial condition as a perturbation of homogeneous steady-state and zero-flux boundary condition to examine oscillatory dynamics in the auxin signalling pathway, Figs.~\ref{fig:aux-osc},~\ref{fig:Middleton_Growth}. 
We solve numerically model \eqref{eq:aux_flux-signal_cell}-\eqref{eq:growth_b} with zero initial condition and zero-flux boundary condition to examine the emergence of the reverse-fountain pattern of auxin distribution at the root tip, Fig.~\ref{fig:Growth2}.
For model \eqref{eq:aux_flux-signal_cell}-\eqref{eq:growth_b}, \eqref{eq:strain} we consider  zero initial condition and zero-flux boundary condition to examine how varying the weighting between flux-induced and strain-induced PIN localisation affects the emergence of the reverse-fountain pattern in auxin distribution at the root tip, Fig.~\ref{fig:Mech}.
We compare solutions of model \eqref{eq:aux_flux-signal_cell}, \eqref{eq:Ap-AUX-cell}, \eqref{eq:Ap-pas}, \eqref{eq:Ap-Fluxes_a}, \eqref{eq:Ap-Fluxes_c}, model \eqref{eq:aux_flux-signal_cell}, \eqref{eq:Ap-AUX-cell}, \eqref{eq:Ap-pas}, \eqref{eq:Ap-Fluxes_a}, \eqref{eq:Ap-Fluxes_d}, model \eqref{eq:aux_flux-signal_cell}, \eqref{eq:Ap-AUX-cell}, \eqref{eq:Ap-pas}, \eqref{eq:Ap-Fluxes_b}, \eqref{eq:Ap-Fluxes_c},  and model \eqref{eq:aux_flux-signal_cell}, \eqref{eq:Ap-AUX-cell}, \eqref{eq:Ap-pas}, \eqref{eq:Ap-Fluxes_b}, \eqref{eq:Ap-Fluxes_d},  completed with periodic boundary conditions and  initial condition as a perturbation of the homogeneous steady-state,   to examine the influence of different mechanisms for auxin transport and PIN localisation on patterns in  auxin  distribution, Fig.~\ref{fig:Apo3indF-kappa}.

Python code used to solve our models numerically is available online \cite{Auxin_Models}.

%----------------------------------------------------------------------------------------------
\subsection*{Pattern Recognition}
%----------------------------------------------------------------------------------------------

For two (neighbouring) cells A and B auxin is defined as flowing from cell A to cell B if $J_{a}^{AB}/\lambda > \theta$.  We define two cells  A and B as being connected if, potentially via some chain of other cells, auxin flows  from A into B and from B into A, through different membranes.   In the absence of any source or sink cells, we define a passage as a set of pairwise connected cells that splits the rest of the domain into two distinct sub-domains, for the three-cell case a passage occupies the entire domain i.e.~each cell is connected to every other cell.  A spot is defined as either a single cell which experiences only auxin influx, or a set of pairwise connected cells that does not split the domain into sub-domains and experiences inflow from at least one neighbouring cell.

%----------------------------------------------------------------------------------------------
\subsection*{ Linear   Stability Analysis}
%----------------------------------------------------------------------------------------------

Linear  stability analysis of model \eqref{eq:aux_flux-signal_cell}-\eqref{eq:aux_flux-signal_fluxes} was performed to determine how changes in parame\-ter values affect the auxin distribution in a plant tissue. The simplified domain considered in the stability analysis is a ring of three cells with each cell having two neighbours and communicating with every other cell in the domain.  The outer boundary, i.e.\ cell borders which do not border other cells, has a zero-flux boundary condition.  This is the smallest domain such that each cell communicates uniquely with every other cell. A similar approach was considered in \cite{Hayakawa_Y_2015}.

We used standard numerical continuation techniques via the \texttt{Matcont} package in \texttt{MATLAB}, \cite{matcont}, to examine the stability of steady-state solutions of model \eqref{eq:aux_flux-signal_cell}-\eqref{eq:aux_flux-signal_fluxes}. This approach was used to determine the stability of the homogeneous steady-state solution of model \eqref{eq:aux_flux-signal_cell}-\eqref{eq:aux_flux-signal_fluxes} in order to locate the regions of the $h-\beta_{p}$ parameter space for which heterogeneous pattern formation is possible. We further used these methods to investigate the possibility of oscillatory solutions of model \eqref{eq:aux_flux-signal_cell}-\eqref{eq:aux_flux-signal_fluxes} upon variation of $\alpha_{m}$ and $\beta_{p}$.

To investigate the likelihood of occurrence of different pattern types when pattern formation can occur, model \eqref{eq:aux_flux-signal_cell}-\eqref{eq:aux_flux-signal_fluxes} was solved for various values of $\beta_{p}$  for different initial conditions  defined as random perturbations of homogeneous steady-state.

%==============================================================================================
\section{Results}
%==============================================================================================

\begin{figure}\centering
 \includegraphics[width=\linewidth]{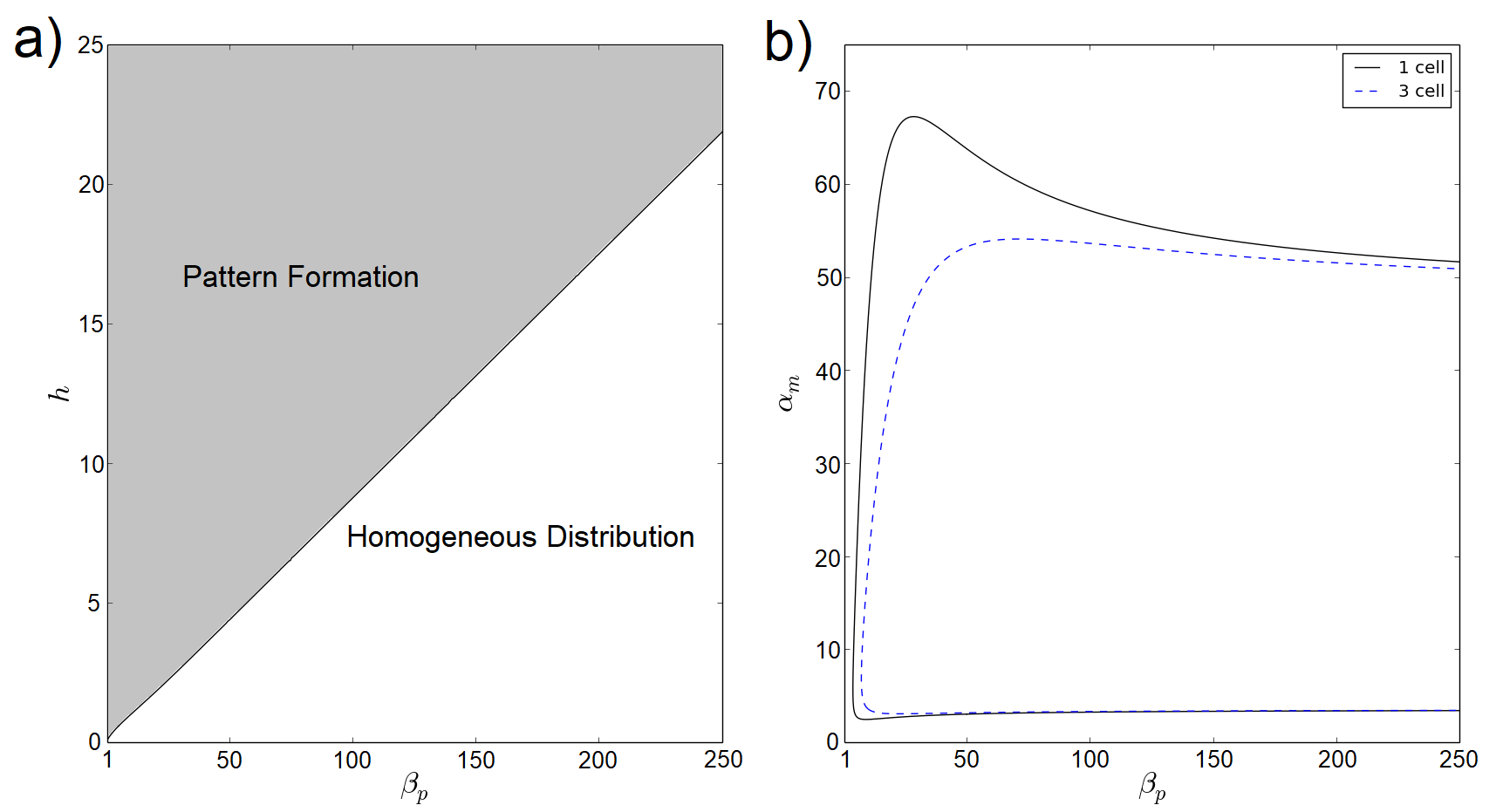}
 \caption{Analysis of parameter inference on solution types of model \eqref{eq:aux_flux-signal_cell}-\eqref{eq:aux_flux-signal_fluxes}. \textbf{a)} the boundary between pattern formation (shaded area) and homogeneous distribution (non-shaded area) is defined by an approximately linear relationship between $h$ and $\beta_{p}$. Minimum value of $h$ presented here is $0.081$. \textbf{b)} as $\alpha_{m}$ and $\beta_{p}$ are varied, model \eqref{eq:aux_flux-signal_cell}-\eqref{eq:aux_flux-signal_fluxes} undergoes a Hopf bifurcation and is able to have oscillatory solutions. For a single cell model, i.e.~$P_{ij} = 0\  \forall\ i,j$, oscillatory solutions occur within the area bounded by the black solid line. For the three-cell geometry used for analysis model \eqref{eq:aux_flux-signal_cell}-\eqref{eq:aux_flux-signal_fluxes} is able to generate oscillatory solutions in the smaller area bounded by the dashed blue line.}
 \label{fig:Summary2}
\end{figure}

\begin{figure}\centering
 \begin{subfigure}[b]{0.48\textwidth}
  \includegraphics[width=\linewidth]{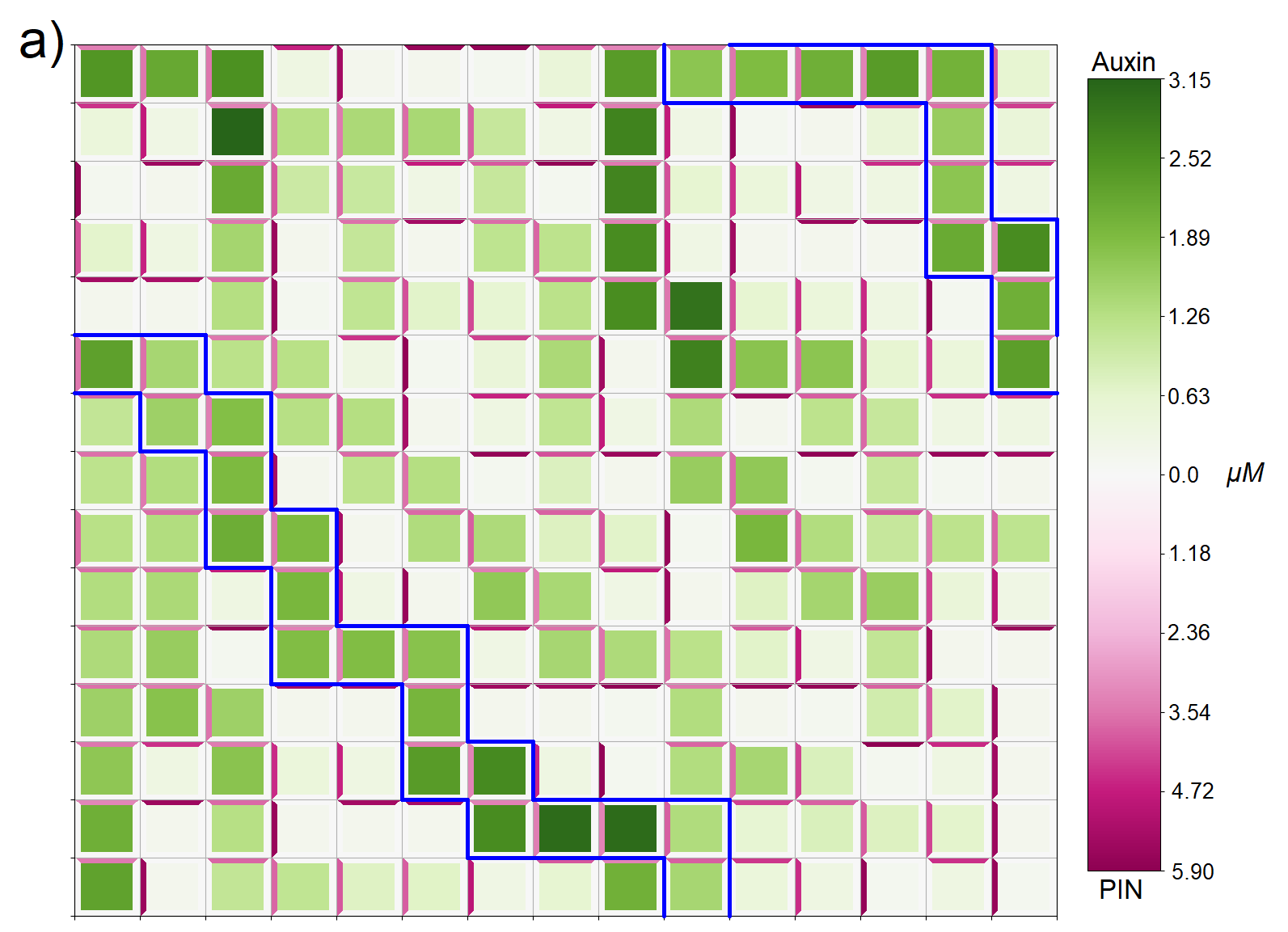}
 \end{subfigure}
 ~
 \begin{subfigure}[b]{0.48\textwidth}
  \includegraphics[width=\linewidth]{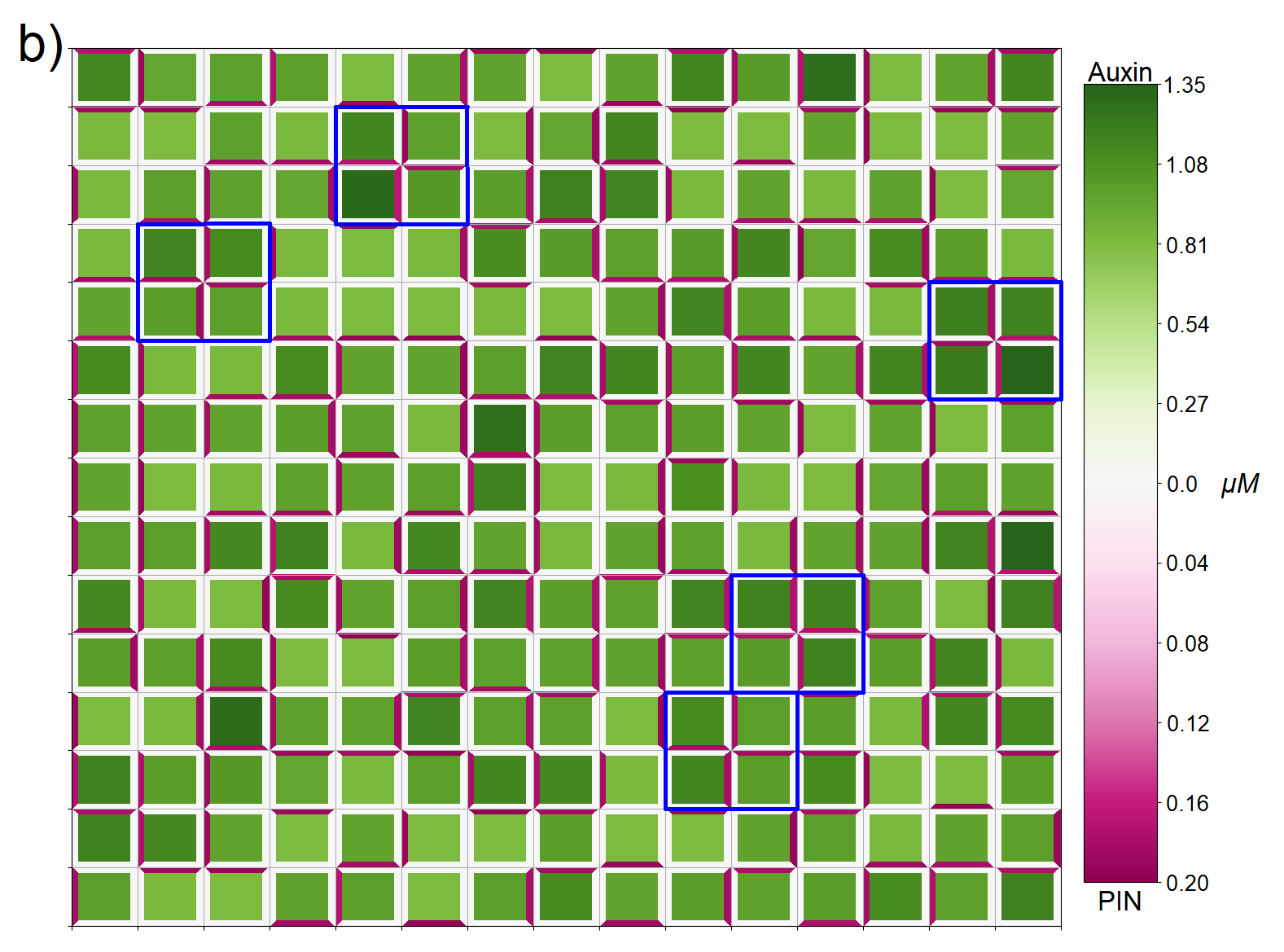}
 \end{subfigure}
 \caption{Mathematical model including auxin signalling can generate both passage- and spot-type patterns of auxin distribution. Numerical solution of model \eqref{eq:aux_flux-signal_cell}-\eqref{eq:aux_flux-signal_fluxes} on a regular lattice of cells with $\lambda=0.5$. Green colour represents the concentration of cellular auxin, darker shades correspond to higher concentrations. Magenta colour represents the concentration of membrane-bound PIN, darker shades correspond to higher concentrations. \textbf{a)} for low values of the rate of PIN binding to auxin-TIR1 $\beta_{p}$,  here $\beta_{p} = 5$, passage patterns of auxin distribution are formed. \textbf{b)} for higher values of $\beta_{p}$, here $\beta_{p}=100$, spot patterns of auxin distribution can be formed, here  blue borders indicate spots. All other parameter values are described in Tables~\ref{tab:signal} and \ref{tab:flux}. Periodic boundary conditions were used for both simulations.}
 \label{fig:Basic_Sig2-Periodic-Spatial-7014}
\end{figure}

Unless otherwise specified, parameter values are taken as the default values listed in Tables~\ref{tab:signal},~\ref{tab:flux}.  To analyse the effect of auxin-related signalling processes on auxin transport and heterogeneous distribution in a plant tissue we considered model \eqref{eq:aux_flux-signal_cell}-\eqref{eq:aux_flux-signal_fluxes}  in the three cell ring domain for a wide range of  parameters corresponding to the coupling between transport and signalling processes, i.e.\ the rate of PIN binding to auxin-TIR1 $\beta_{p}$, and the sensitivity of the flux-feedback function $h$. 
The stability analysis shows that increase in sensitivity of auxin-induced PIN degradation to auxin  ($\beta_{p}$ increases) leads to transition of stable heterogeneous solutions   to homogeneous steady-states, Fig.~\ref{fig:Summary2}a).
For sufficiently high values of $h$ and appropriate values of $\beta_{p}$, both spot and passage type patterns of auxin distribution are possible.
Although both types of patterns were obtainable in the parameter region indicated in Fig.~\ref{fig:Summary2}a), the  value of $\beta_{p}$ has a great influence on the probability of each pattern emerging. To test the influence of parameters on probability of the emergence of specific pattern types we used a lattice of 3x3 cells with periodic boundary condition.
Specifically, we found that as $\beta_{p}$ increased the ratio of occurrences of passage patterns to spot patterns reduced dramatically, from as high as $\approx 95\%$ for $\beta_{p} = 1$ to $\approx25\%$ for $\beta_{p} = 100$, with fixed value $h = 50$.
The type of boundary conditions also has an effect on the pattern formation  for model \eqref{eq:aux_flux-signal_cell}-\eqref{eq:aux_flux-signal_fluxes}.
For zero-flux boundary conditions the probability of the emergence of passage patterns was $\approx 40\%$ for $\beta_{p} = 1$,  and $\approx 20\%$ for $\beta_{p} = 100$.
Since the value of $h$ determines the range of values of $\beta_{p}$ for which heterogeneous patterns can emerge the probability of certain pattern types emerging for specific values of $\beta_{p}$ will also vary with $h$.
Although $\beta_{a}$ and $\beta_{p}$ have similar roles in determining the degradation rate of PIN since they influence the rate of PIN-auxin-TIR1 binding, we found that upon varying $\beta_{a}$ from $0.5$ to $50$ for fixed values of $\beta_{p}$ between $1$ and $250$ the probability of emergence of specific pattern types was unchanged, with fixed value $h = 50$.
Numerical simulation results for the cases of both passage and spot patterns in auxin distribution are included in Fig.~\ref{fig:Basic_Sig2-Periodic-Spatial-7014} for $\lambda = 0.5$.
 We tested a small set of values of $\lambda$, and for higher values of   $6\geq\lambda\gtrsim 1.5$  the concentration of membrane-bound PIN is increased leading to much higher concentrations of auxin in spot and passage cells compared to other cells and thus very well defined patterns, whereas for $0 < \lambda\lesssim 1.5$ membrane-bound PIN levels resemble those in Fig.~\ref{fig:Basic_Sig2-Periodic-Spatial-7014} (data not shown).
When relating model \eqref{eq:aux_flux-signal_cell}-\eqref{eq:aux_flux-signal_fluxes} to  symplast-apoplast model  \eqref{eq:aux_flux-signal_cell},  \eqref{eq:Ap-AUX-cell}-\eqref{eq:Ap-Fluxes}  and considering in  \eqref{eq:aux_flux-signal_fluxes}  concentration-induced PIN localisation as in \eqref{eq:Ap-Fluxes_c}, spots formed of two cells that have strong PIN alignment between them emerge using parameter values that would lead to passages for \eqref{eq:aux_flux-signal_cell}-\eqref{eq:aux_flux-signal_fluxes} (data not shown).

\begin{figure}\centering 
 \begin{subfigure}[b]{\textwidth}
  \includegraphics[width=\linewidth]{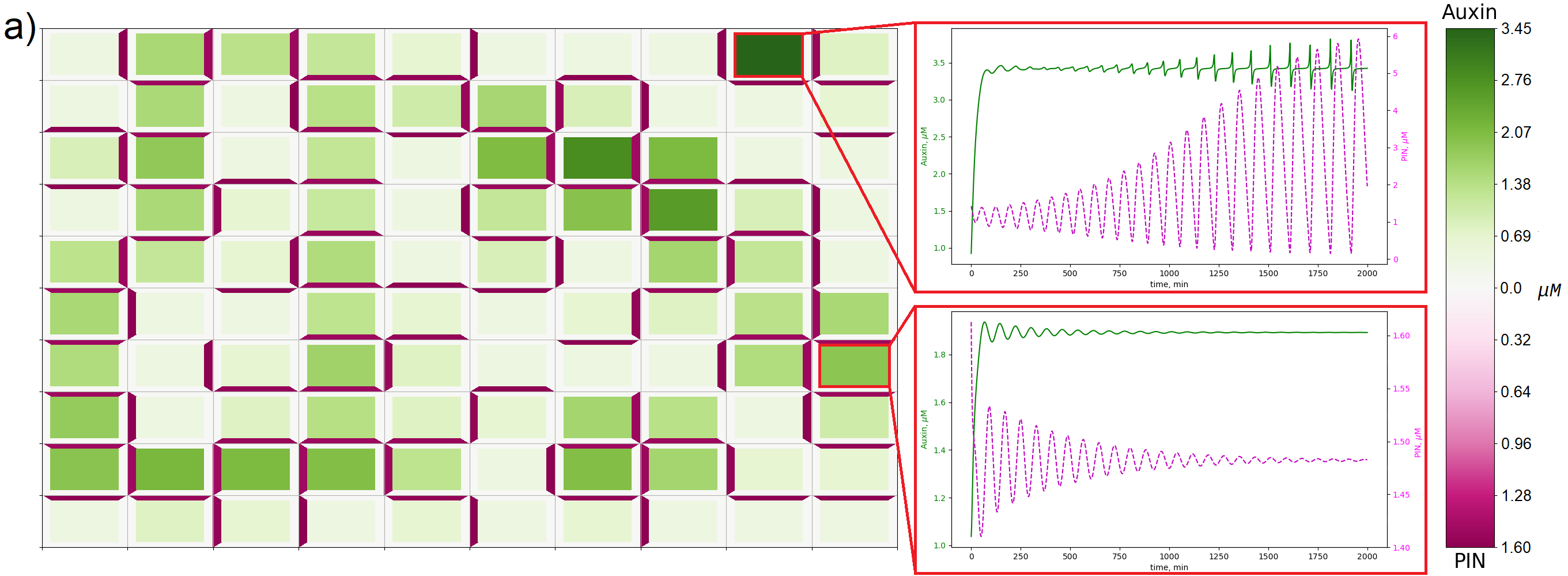}
 \end{subfigure}
 \\
 \begin{subfigure}[b]{\textwidth}
  \includegraphics[width=\linewidth]{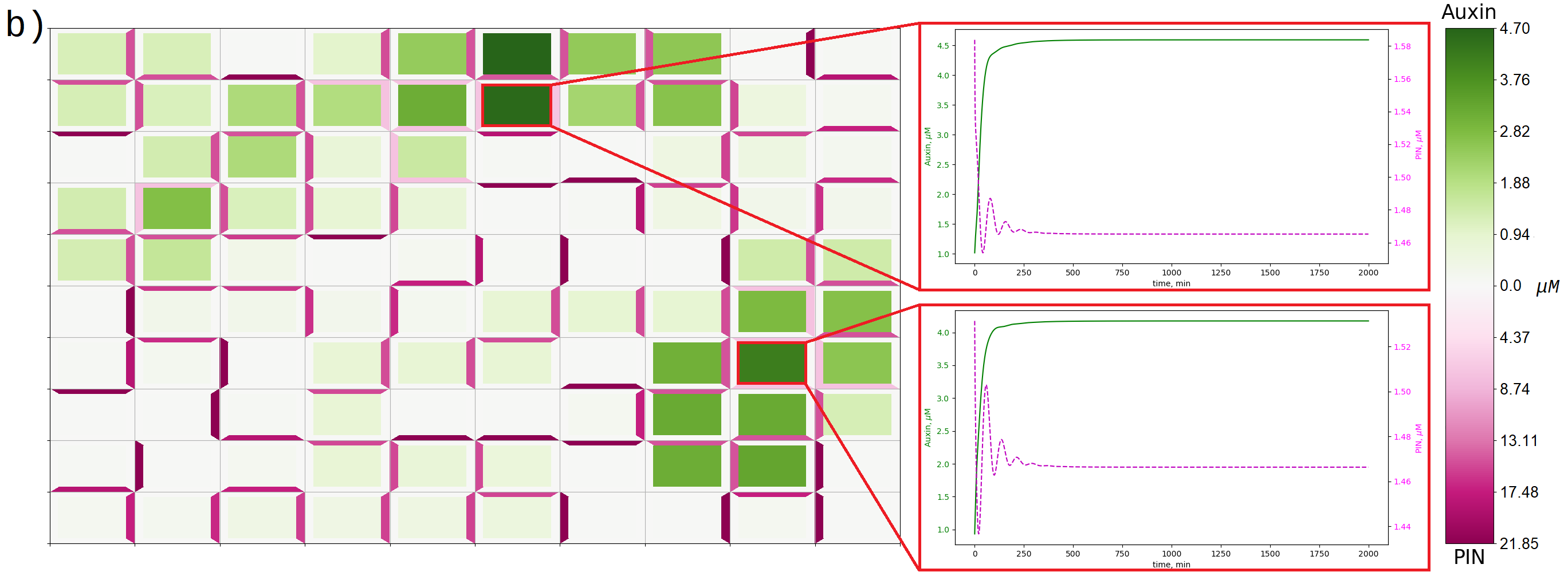}
 \end{subfigure}
 \caption{Oscillatory dynamics in the components of the auxin signalling pathway for appropriate rates of PIN membrane localisation. Numerical solution of model \eqref{eq:aux_flux-signal_cell}-\eqref{eq:aux_flux-signal_fluxes} on a regular lattice of cells, with zero-flux boundary conditions, and a parameter set listed in Table \ref{tab:signal} that generates oscillatory dynamics in a single cell model for auxin-related signalling pathway, \textbf{a)} for $\lambda = 0.05$ PIN concentration is reduced to within physically realistic ranges and single spot cells with oscillating concentrations of auxin and cytoplasmic PIN are generated. \textbf{b)} for $\lambda = 0.5$ the concentration of PIN on cell membranes rises above physically realistic ranges  of $\approx 0-5 \ \mu$M and numerical solutions display no oscillatory dynamics. }
 \label{fig:aux-osc}
\end{figure}

\begin{figure}\centering
 \begin{subfigure}[b]{\textwidth}
  \includegraphics[width=\linewidth]{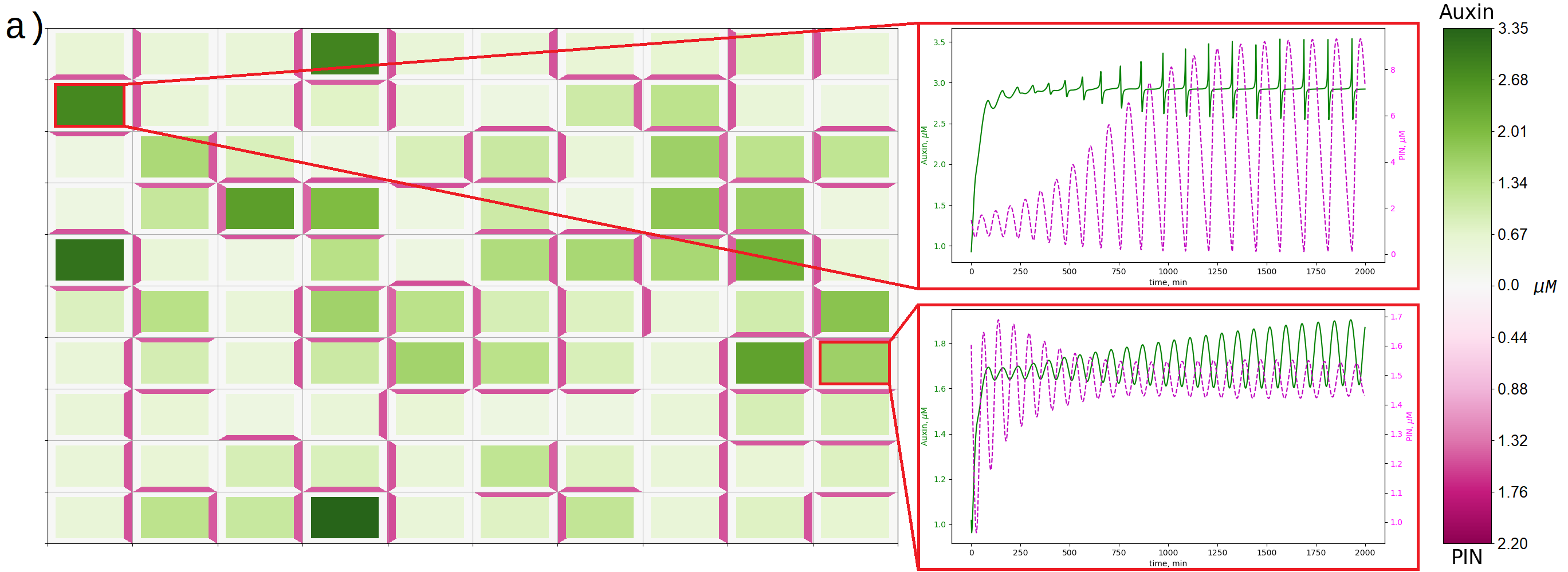}
 \end{subfigure}
 \\
 \begin{subfigure}[b]{\textwidth}
  \includegraphics[width=\linewidth]{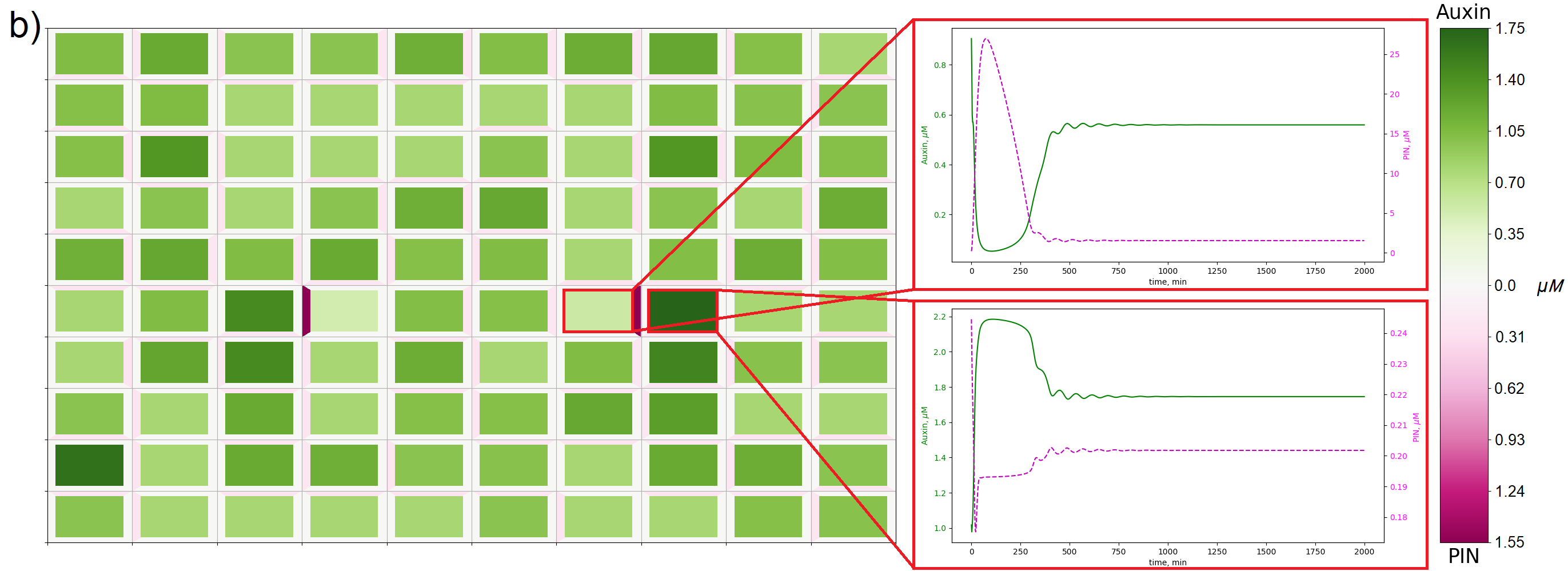}
 \end{subfigure}
 \caption{Oscillatory dynamics are tissue-dependent and robust to growth. \textbf{a)} numerical solution of model \eqref{eq:aux_flux-signal_cell}-\eqref{eq:growth_a} on a regular lattice of growing cells with oscillatory parameters, see Tables~\ref{tab:signal},\ref{tab:flux}, and zero-flux boundary conditions. Oscillatory single-cell spots are now joined by four-cell spots which also have oscillatory dynamics. \textbf{b)} numerical solution of model \eqref{eq:aux_flux-signal_cell}-\eqref{eq:aux_flux-signal_fluxes} on a regular lattice of cells, with zero-flux boundary conditions. Oscillatory parameters have been set for cells (5,4) and (5,7), all other cells have standard parameter values, see Tables~\ref{tab:signal} and \ref{tab:flux}. For modified oscillatory cells, oscillatory dynamics are not preserved.}
 \label{fig:Middleton_Growth}
\end{figure}

Oscillations in  the concentrations of targets of auxin-responsive ARFs have been observed experimentally in the protoxylem cells in the root meristem \cite{DeSmet_I_2007}. Considering the mathematical model for the auxin-related signalling pathway in a single plant cell, it has been shown in \cite{Middleton_A_2010} that for certain  parameter values solutions of the mathematical model can exhibit oscillatory dynamics. Here we  demonstrated that our modified auxin signalling model \eqref{eq:aux_flux-signal_cell}-\eqref{eq:aux_flux-signal_fluxes} can have oscillatory solutions in the case  when considering the dynamics in a single cell (with zero fluxes between cells) and in the case  of PIN-mediated auxin transport  in the three-cell domain, Fig.~\ref{fig:Summary2}b).
To analyse the effect of auxin transport on the oscillatory behaviour of  component of the auxin signalling pathway inside cells in a tissue, we consider model \eqref{eq:aux_flux-signal_cell}-\eqref{eq:aux_flux-signal_fluxes} with the set of parameters for which oscillations in auxin concentration in the single cell model would occur.
We found that when considering oscillatory set of parameters for all cells in a tissue and reducing the rate of PIN localisation to the membrane by a factor of 10, i.e.~$\lambda = 0.05$, we obtain spot-type patterns in  auxin distribution with some spot cells constituting oscillations in  the levels of components of the auxin signalling pathway,  which was not possible for the previously considered value of $\lambda = 0.5$, Fig.~\ref{fig:aux-osc}.   We tested a small set of values of $0 < \lambda \leq 0.5$, and observed oscillations in numerical simulations for $0 < \lambda \lesssim 0.35$ but did not observe oscillations for $0.35\lesssim \lambda \leq 0.5$.
Interestingly, single cell type spots demonstrate oscillatory dynamics in auxin concentration, whereas four-cells-size spots do not present oscillations in the auxin concentration. 
When considering oscillatory parameters alongside  horizontal growth however, \eqref{eq:aux_flux-signal_cell}-\eqref{eq:growth_a},  not only are single spot cells with auxin concentration oscillating around a high value generated, but oscillatory dynamics are present in four-cells-size spots with smaller period, Fig.~\ref{fig:Middleton_Growth}a).
Assuming that different cells have different properties of the auxin signalling pathway, in two cells we considered the set of model parameters that would lead to oscillatory dynamics, with all other cells having standard parameter values, see Table~\ref{tab:signal}. In this case, no stable oscillations in the auxin dynamics were observed and as the steady-state distribution of auxin we have low auxin concentrations in the two modified cells and high auxin concentration in the neighbouring cells which experience strong auxin flux from the modified cells, and all other cells have low concentrations of membrane-bound PIN ($\approx 0.25 \ \mu$M), Fig.~\ref{fig:Middleton_Growth}b).
We further considered model parameters that would lead to oscillatory dynamics, i.e. as in Fig.~\ref{fig:aux-osc}, in a single cell model in all cells within a specified radius from the central cell and standard parameter values, i.e. as in Fig.~\ref{fig:Basic_Sig2-Periodic-Spatial-7014},  in all other cells.
In each case, apart from when every cell had parameters that would lead to oscillatory dynamics in a single cell model, the oscillatory dynamics in auxin concentration were not persistent (data not shown).
This suggests that in order to generate oscillations in the  levels of targets of ARF observed in protoxylem cells, changes in the signalling pathways in all cells of a plant tissue are required.

\begin{figure}\centering
 \begin{subfigure}[b]{0.48\textwidth}
  \includegraphics[width=\linewidth]{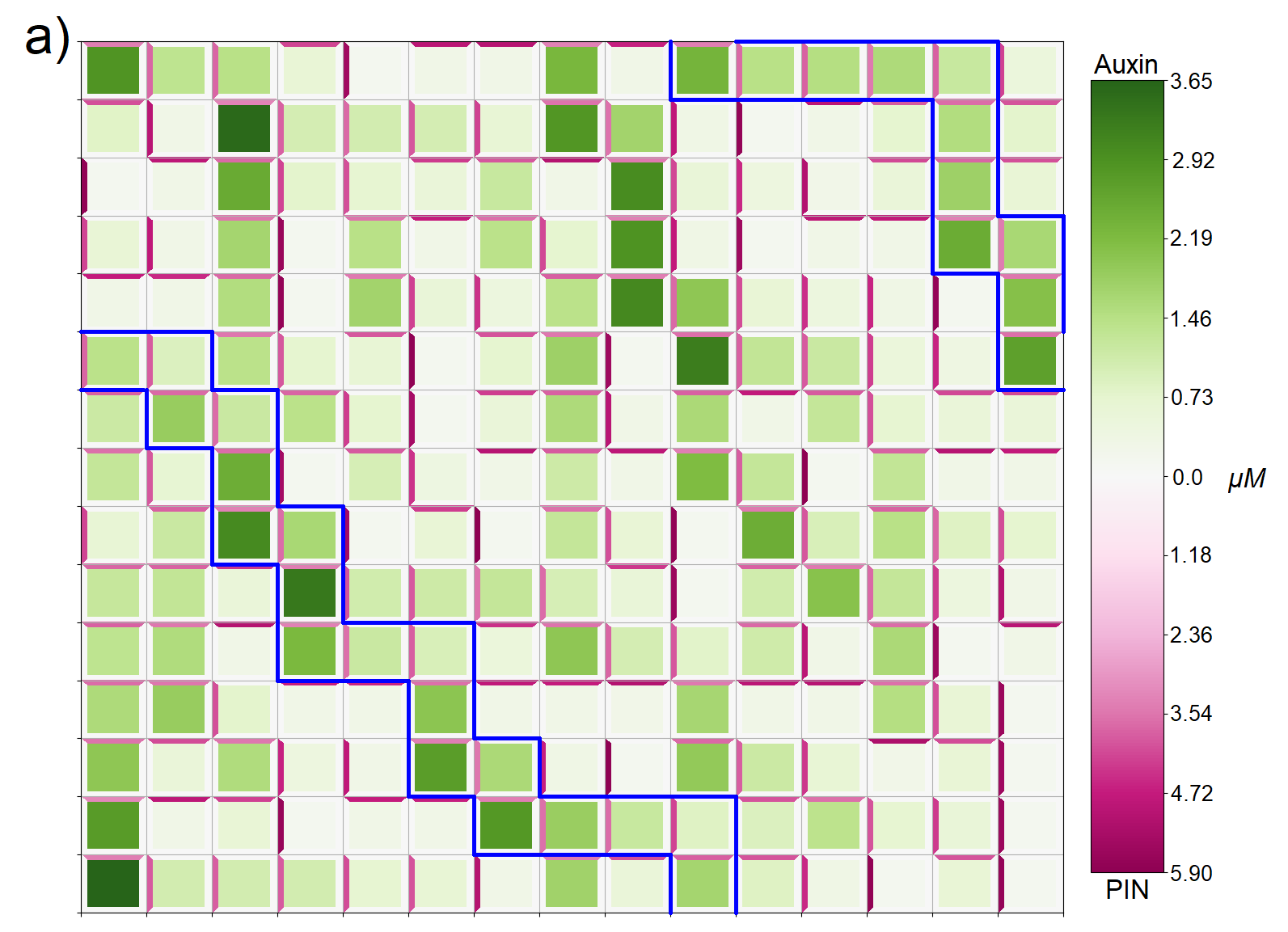}
 \end{subfigure}
 ~
 \begin{subfigure}[b]{0.48\textwidth}
  \includegraphics[width=\linewidth]{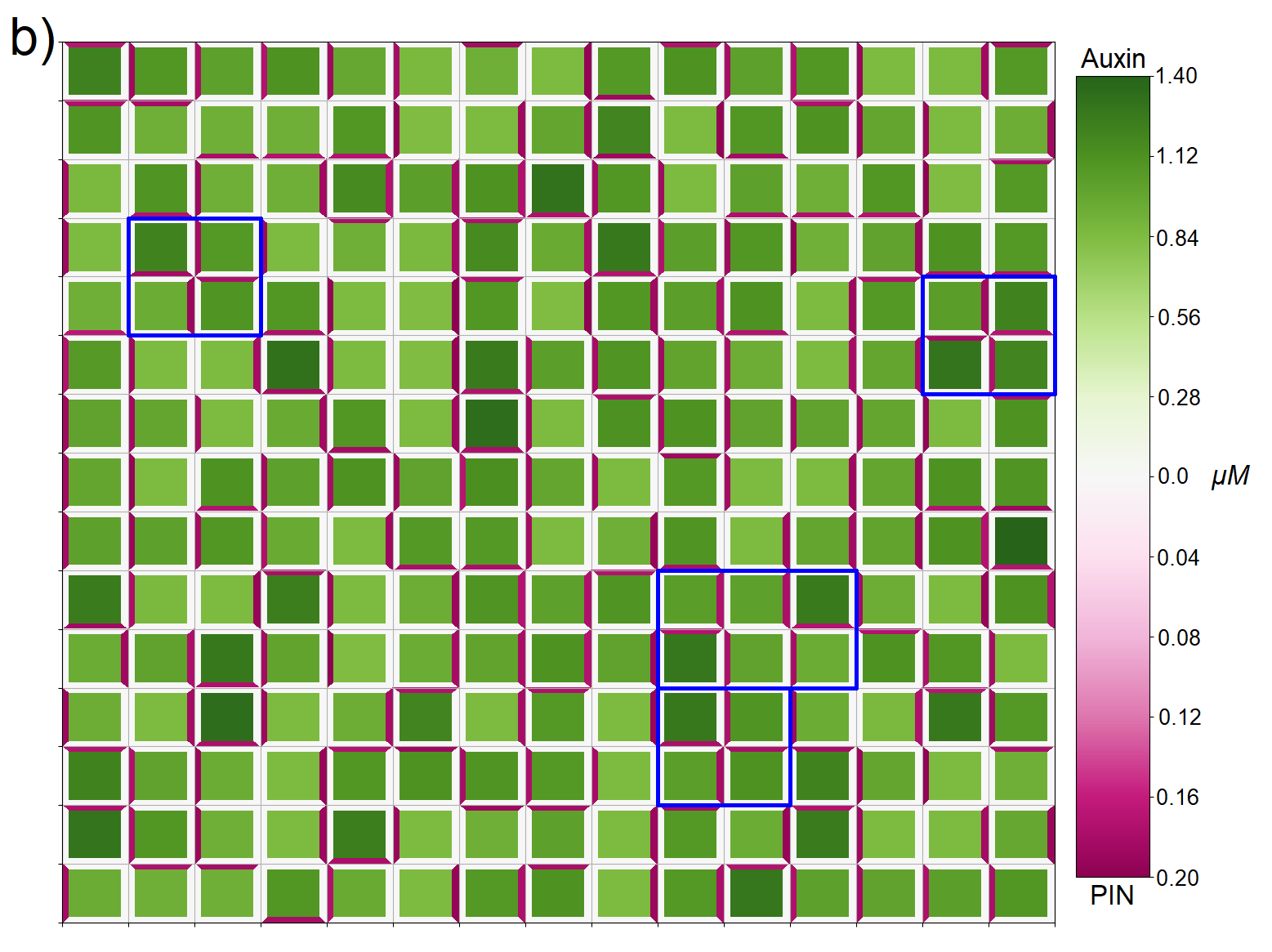}
 \end{subfigure}
 \caption{PIN polarisation aligns with oriented cell growth. Numerical solution of model \eqref{eq:aux_flux-signal_cell}-\eqref{eq:growth_a} on a regular lattice of cells, starting from the same initial conditions as in Fig.~\ref{fig:Basic_Sig2-Periodic-Spatial-7014}, but with cells undergoing auxin-dependent, horizontal growth. \textbf{a)} In the passage parameter regime oriented growth has no effect on the placement of cells within the passage, and only shifts the PIN alignment   from vertical to horizontal for five cells. \textbf{b)} In the spot parameter regime oriented growth disturbs the formation of spots, halting the emergence of one and enlarging another, and shifts the PIN alignment   from vertical to horizontal for 36 cells. All parameters are described in Tables~\ref{tab:signal} and \ref{tab:flux}. Cells are represented in the reference configuration.}
 \label{fig:Sig_horizontal}
\end{figure}

\begin{figure}\centering
 \begin{subfigure}[b]{0.48\textwidth}
  \includegraphics[width=\linewidth]{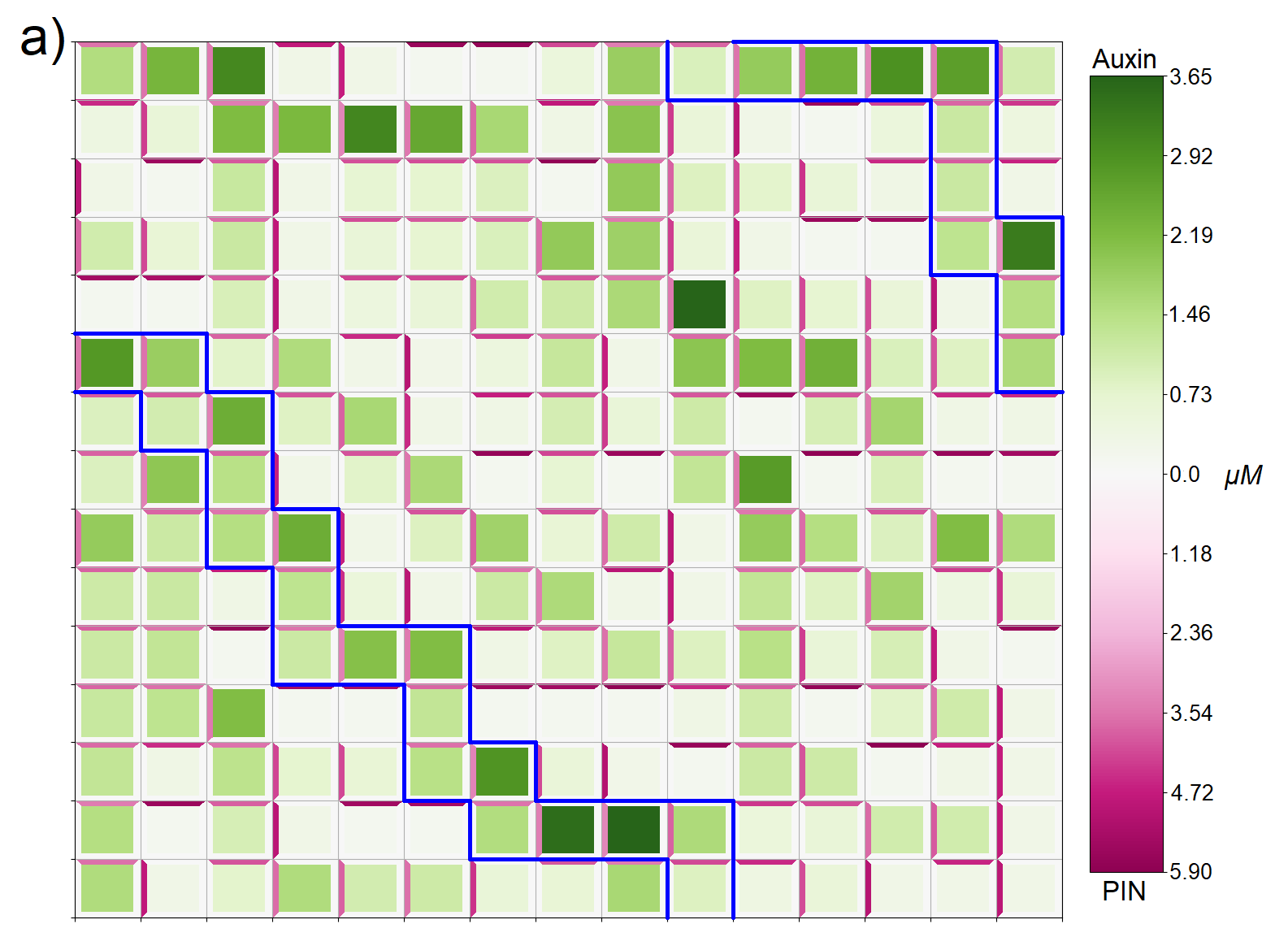}
 \end{subfigure}
 ~
 \begin{subfigure}[b]{0.48\textwidth}
  \includegraphics[width=\linewidth]{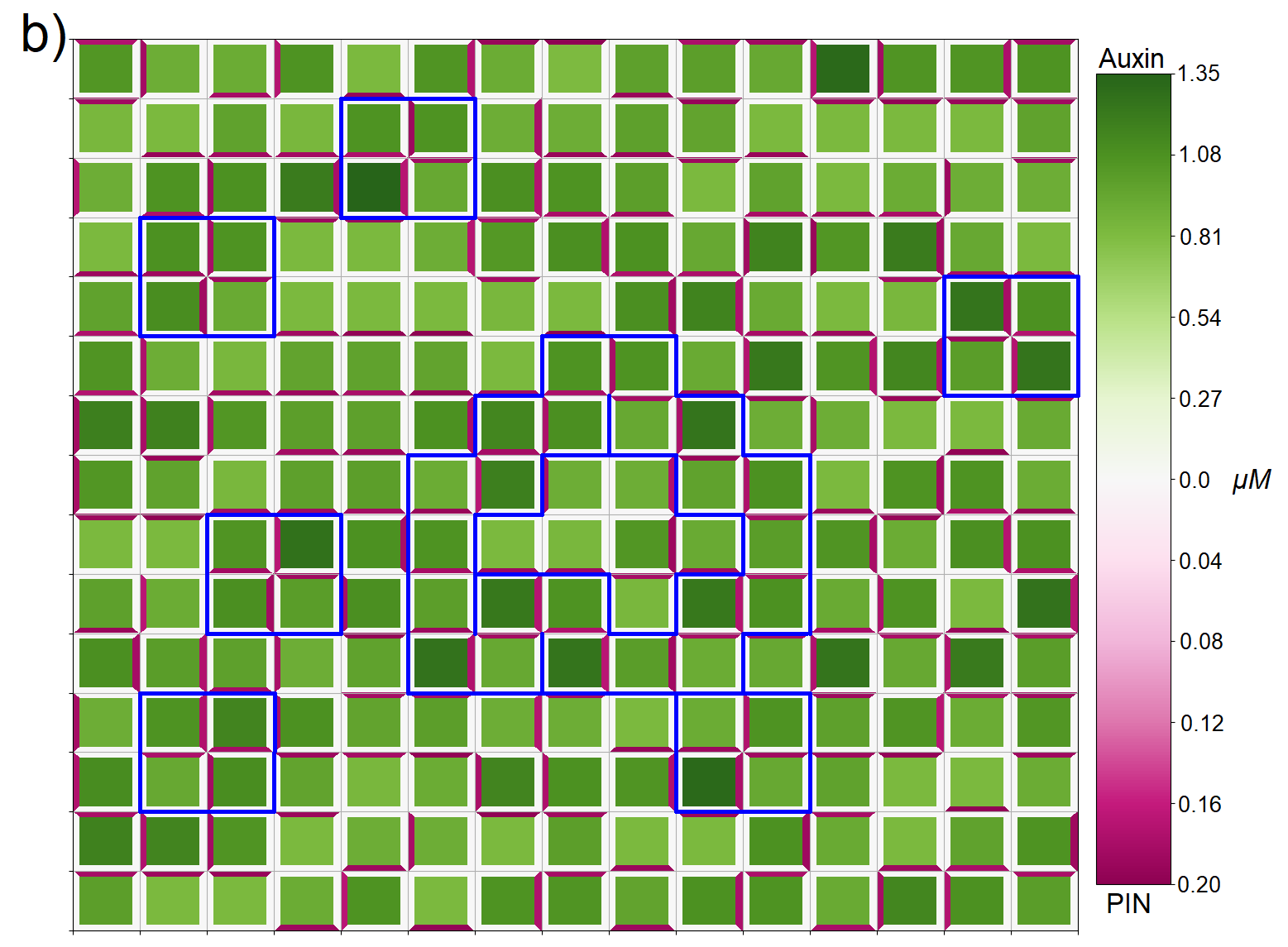}
 \end{subfigure}
 \caption{PIN polarisation aligns with oriented cell growth. Numerical solution of model \eqref{eq:aux_flux-signal_cell}-\eqref{eq:growth_a} on a regular lattice of cells, starting from the same initial conditions as in Fig.~\ref{fig:Basic_Sig2-Periodic-Spatial-7014}, but with cells undergoing auxin-dependent, vertical growth. \textbf{a)} In the passage parameter regime oriented cell growth has no effect on the placement of cells within the passage, and only shifts the PIN alignment   from horizontal to vertical for four cells. \textbf{b)} In the spot parameter regime oriented growth disturbs the formation of spots,  leading to the emergence of two new spots and modifying another into a small passage, and shifts the PIN alignment  from horizontal to vertical for 35 cells.  All parameters are described in Tables~\ref{tab:signal} and \ref{tab:flux}. Cells are represented in the reference configuration.}
 \label{fig:Sig_vertical}
\end{figure}

\begin{figure}\centering
 \begin{subfigure}[b]{0.48\textwidth}
  \includegraphics[width=\linewidth]{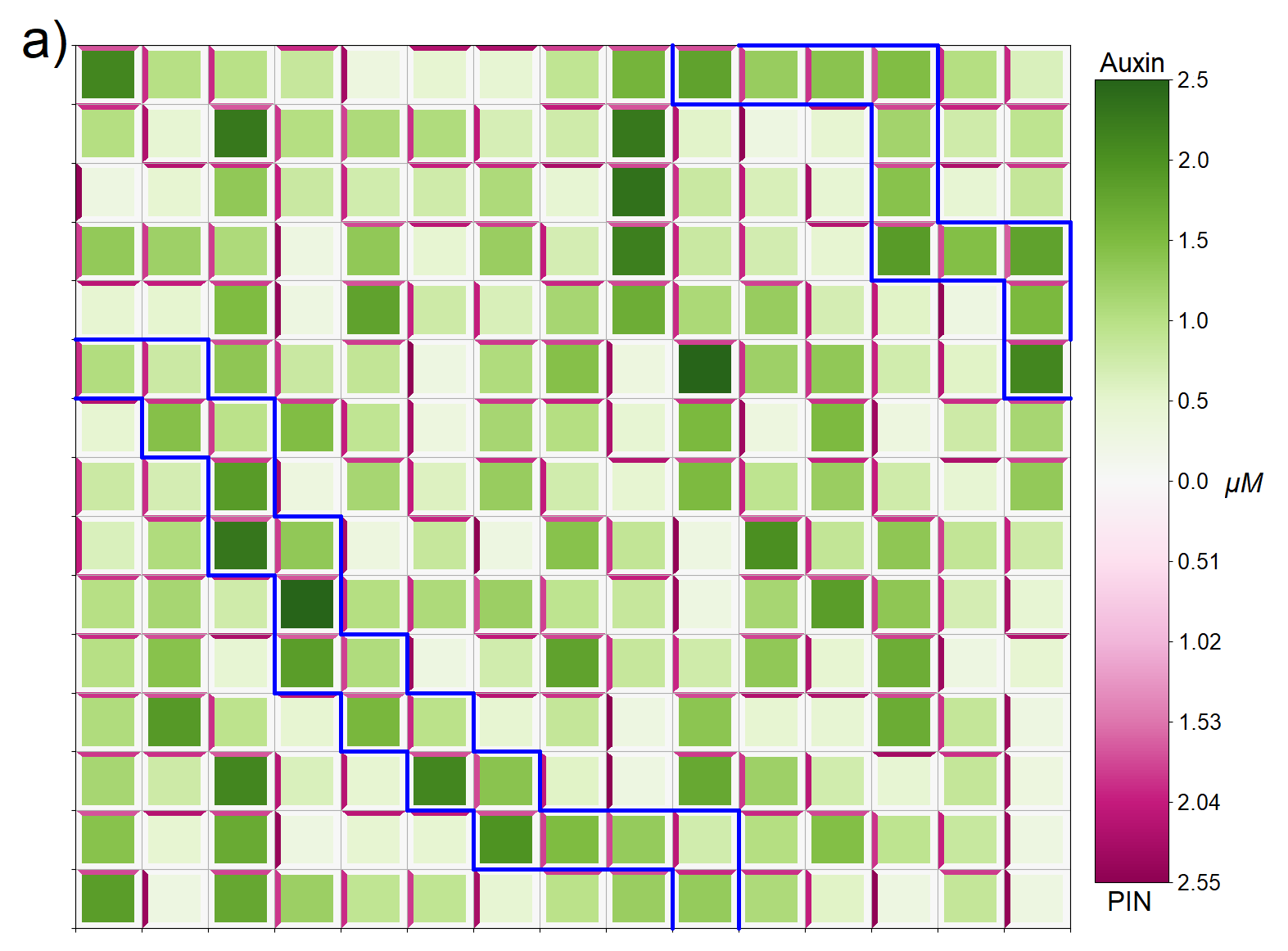}
 \end{subfigure}
 ~
 \begin{subfigure}[b]{0.48\textwidth}
  \includegraphics[width=\linewidth]{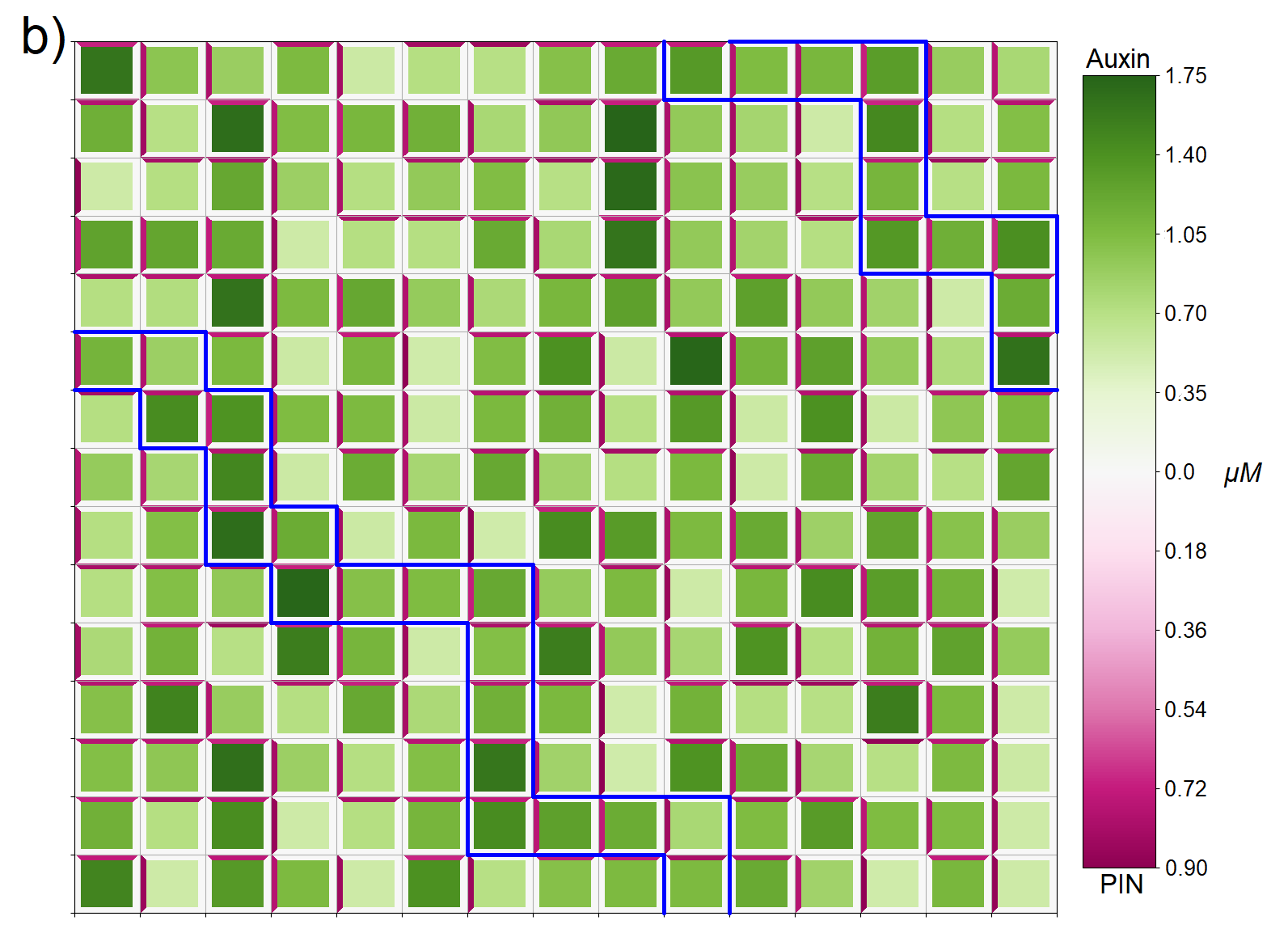}
 \end{subfigure}
 \caption{Relative weighting of chemical and mechanical feedback on PIN localisation. Numerical solution of model \eqref{eq:aux_flux-signal_cell}-\eqref{eq:growth_a},\eqref{eq:strain} on a regular lattice of cells, starting from the same initial conditions as in Fig.~\ref{fig:Basic_Sig2-Periodic-Spatial-7014}a), but with cells undergoing auxin-dependent, horizontal growth, and with strain-dependent PIN localisation. \textbf{a)} when chemical and mechanical PIN localisation are weighted equally, $\lambda=\nu=0.25$, four passage cells undergo small shifts and a total of 10 cells have altered PIN alignment, with one cell shifting from horizontal to vertical, and nine cells shifting from vertical to horizontal. \textbf{b)} When chemical PIN localisation is dominated by mechanical PIN localisation, $\lambda=0.1$ and $\nu=0.4$, eight passage cells undergo small shifts and a total of 21 cells have altered PIN alignment, with three cells shifting from horizontal to vertical, and eighteen cells shifting from vertical to horizontal.   All parameters are described in Tables~\ref{tab:signal} and \ref{tab:flux}. Cells are represented in the reference configuration.}
 \label{fig:Sig_mech}
\end{figure}

The precise role of growth and its influence on the distribution of auxin in plant tissues is still open \cite{Korver_R_2018}. Incorporating the tissue growth in the auxin transport model \eqref{eq:aux_flux-signal_cell}-\eqref{eq:growth_a}, where membrane-bound PIN is diluted on the growing membranes, we found that oriented, either horizontal or vertical,  growth influences the overall PIN polarisation across the growing tissue, i.e.~PIN preferentially polarises along the axis of growth, however this effect does not  alter the type of patterns in the auxin distribution in a tissue (i.e.~passage or spot) but can alter its distribution, Figs.~\ref{fig:Sig_horizontal},~\ref{fig:Sig_vertical}. Interestingly, growth appeared to exert a stronger influence on the polarisation of PIN when considering parameter values that lead to the emergence of spot patterns in the absence of growth.
We found similar results when considering  horizontal and vertical growth simultaneously for identical parameters as in Figs.~\ref{fig:Sig_horizontal},~\ref{fig:Sig_vertical} (data not shown).
It seems that maximal growth rate has limited influence on the overall pattern formation; when $\chi$ was varied between $0.1$ and $10$ there were small changes in the exact concentrations, $\approx 0.3 \ \mu$M for auxin and $\approx 0.1 \ \mu$M for PIN, however PIN polarisation patterns were unchanged.
Incorporating strain-dependent localisation of PIN, see equation~\eqref{eq:aux_flux-signal_cell}-\eqref{eq:growth_a},\eqref{eq:strain}, for equal or stronger weighting of flux-induced compared to strain-induced PIN localisation ($\lambda\geq\nu$) we obtained similar patterns in auxin distribution, Figs.~\ref{fig:Sig_horizontal}a)~and~\ref{fig:Sig_mech}a). However when strain-induced localisation strongly dominates flux-induced localisation ($\lambda <\nu$) a significant reduction in concentrations of both auxin and PIN, compared to the cases where $\lambda\geq\nu$, is observed, Fig.~\ref{fig:Sig_mech}.   To ensure that maximum amount of PIN localised to a cell membrane is consistent with previous simulations we considered a range of values of  $\lambda$ and $\nu$  such that $\lambda + \nu = 0.5$.  For $\lambda\ll\nu$ heterogeneous patterns do not form.

\begin{figure}\centering
 \includegraphics[width=0.9\linewidth]{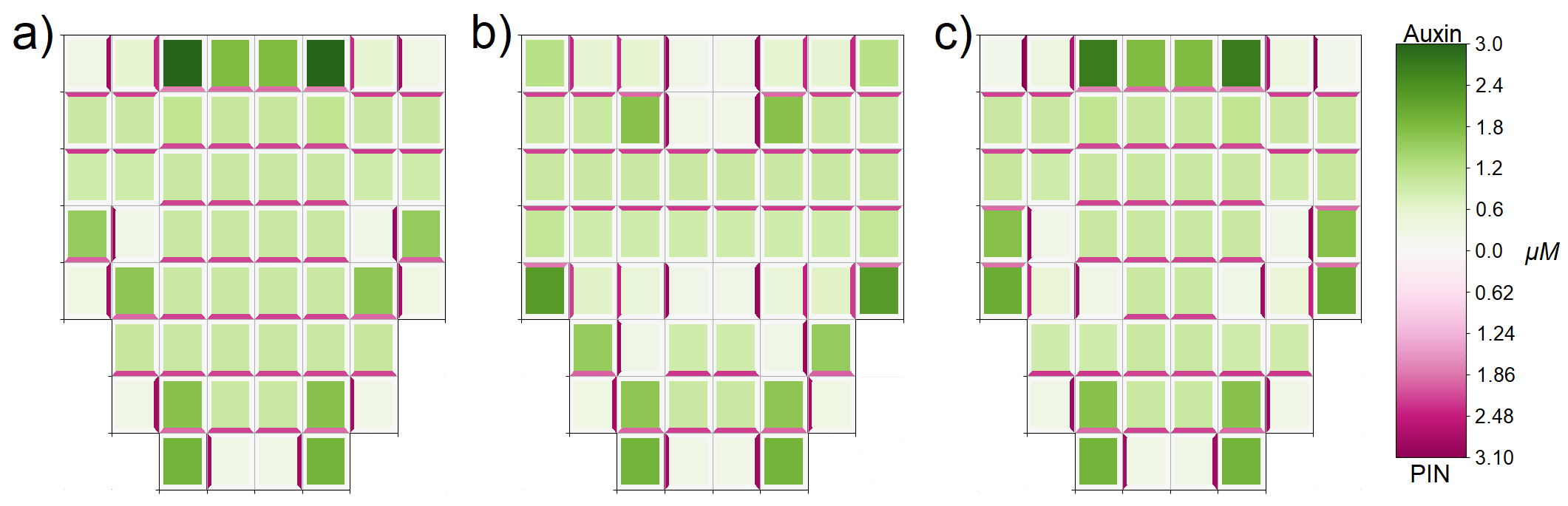}
 \caption{Influence of tissue growth on PIN polarisation contributes to the formation of `reverse-fountain' auxin distribution patterns at the root tip.  Numerical solution of model \eqref{eq:aux_flux-signal_cell}-\eqref{eq:growth_b} on a modified domain and different combinations of source and sink cells. \textbf{a)} the central four cells in the top row are source cells, auxin flows from these cells to the base of the tissue with no flow  from the base cells back up the tissue. \textbf{b)} the outer four cells in the top row are sink cells, auxin flows from cells in the top five rows into these sinks, in the bottom three rows of cells auxin flows to the base of the tissue. \textbf{c)} the central four cells in the top row are source cells and the outer four cells in the top row are sink cells, auxin flows from the source cells down the tissue, the central columns flow to the base of the tissue and the outer columns divert outwards to flow back up to the sink cells, resembling the reverse fountain pattern observed at the root tip. Model parameters are described in Tables~\ref{tab:signal} and \ref{tab:flux}, with zero-flux boundary conditions. Cells are represented in the reference configuration.}
 \label{fig:Growth2}
\end{figure}

To analyse the effect of auxin-related signalling pathway and growth on auxin flux in the plant root tip, we consider model \eqref{eq:aux_flux-signal_cell}-\eqref{eq:growth_b} on a modified lattice of cells resembling a root tip. We assume that growth (cell elongation) occurs only along one axis, i.e.~down the root, and is inhibited by high auxin concentrations and constrained by tissue tension. For consistency with auxin availability in the root, the bulk flow of auxin from shoot to root through the vascular bundle was simulated by including a source term in the central four of the top row of cells. In some simulations we also included sinks in the epidermis cells, outer two cells on each side on the top row, since it is assumed that some auxin is evacuated from the root tip along the epidermis \cite{Swarup_R_2005}. The steady state solutions of model equations \eqref{eq:aux_flux-signal_cell}-\eqref{eq:growth_b}, considering  zero initial conditions and no strain-dependent PIN localisation ($\nu =0$) are presented in Fig.~\ref{fig:Growth2}.
When there are source cells only, auxin flows from the apex of the tissue to the base where it settles, Fig.~\ref{fig:Growth2}a).
When there are sink cells only auxin flows from cells in the top five rows of cells to the sinks, and auxin in the bottom three rows of cells pools at the base of the tissue, Fig.~\ref{fig:Growth2}b).
When both source and sink cells are included a reverse-fountain pattern similar to those observed at the root tip emerges where auxin flows down the root and both pools at the tip and branches out to flow back up the outer layers of cells, Fig~\ref{fig:Growth2}c).

\begin{figure}\centering
 \includegraphics[width=.66\linewidth]{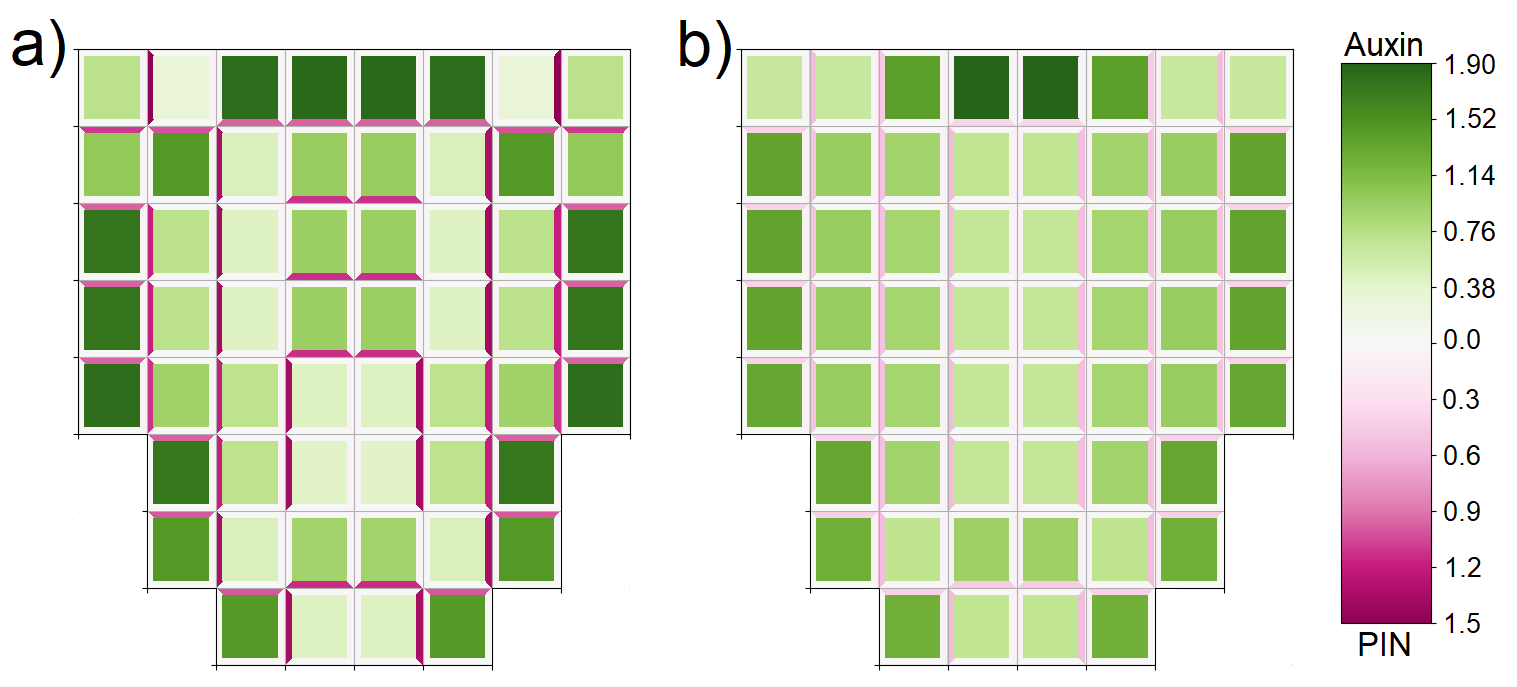}
 \caption{Strain-induced PIN localisation does not significantly effect the formation of reverse flows when weighted below chemically-induced localisation. Model \eqref{eq:aux_flux-signal_cell}-\eqref{eq:growth_b},\eqref{eq:strain} solved on a modified lattice of growing cells, with source and sink cells as in Fig.~\ref{fig:Growth2}. \textbf{a)} strain-induced PIN localisation is weighted equally with flux-induced PIN localisation, $\lambda = \nu = 0.25$. Auxin flows from the source cells to halfway down the tissue where it branches out and then flows back up the outer cells. Auxin produced at the root tip also flows up the outer layer cells. \textbf{b)} strain-induced PIN localisation is weighted above flux-induced PIN localisation, $\lambda = 0.1$, $\nu = 0.4$. Auxin flows directly from source cells to sink cells. Auxin produced in the central file of cells below the source cells does not flow to the base of the tissue, instead immediately flowing outwards to the outer cells where it then flows upwards to the sink cells. All parameters are described in Tables~\ref{tab:signal} and \ref{tab:flux}, with zero-flux boundary conditions. Cells are represented in the reference configuration.}
 \label{fig:Mech}
\end{figure}

To analyse the effect of strain-induced PIN localisation to cell membranes on auxin flux in a plant root tip and its reverse flow  we considered equation \eqref{eq:strain} for a range of values of parameters $\lambda$ (rate of chemical localisation) and $\nu$ (rate of mechanical localisation) such that $\lambda + \nu = 0.5$ so that maximum amount of PIN localised to a cell membrane is consistent with previous simulations. For $\nu < \lambda$ reverse flow patterns may still be generated, when $0.4<\lambda\leq 0.5$ auxin still flows from source cells to the base of the tissue as well as branching out and back up the outer cells, when $0.15\leq\lambda<0.4$ auxin flows only partway down the tissue from the source cells before branching out to flow back up the outer cells and does not flow to the base of the tissue, Fig.~\ref{fig:Mech}a). For $\lambda < 0.15$ formation of a reverse flow pattern was completely inhibited, with auxin flowing directly from source cells to sink cells, and auxin flowing from the base of the tissue to the sinks, Fig.~\ref{fig:Mech}b).
 
\begin{figure}\centering
 \includegraphics[width=\linewidth]{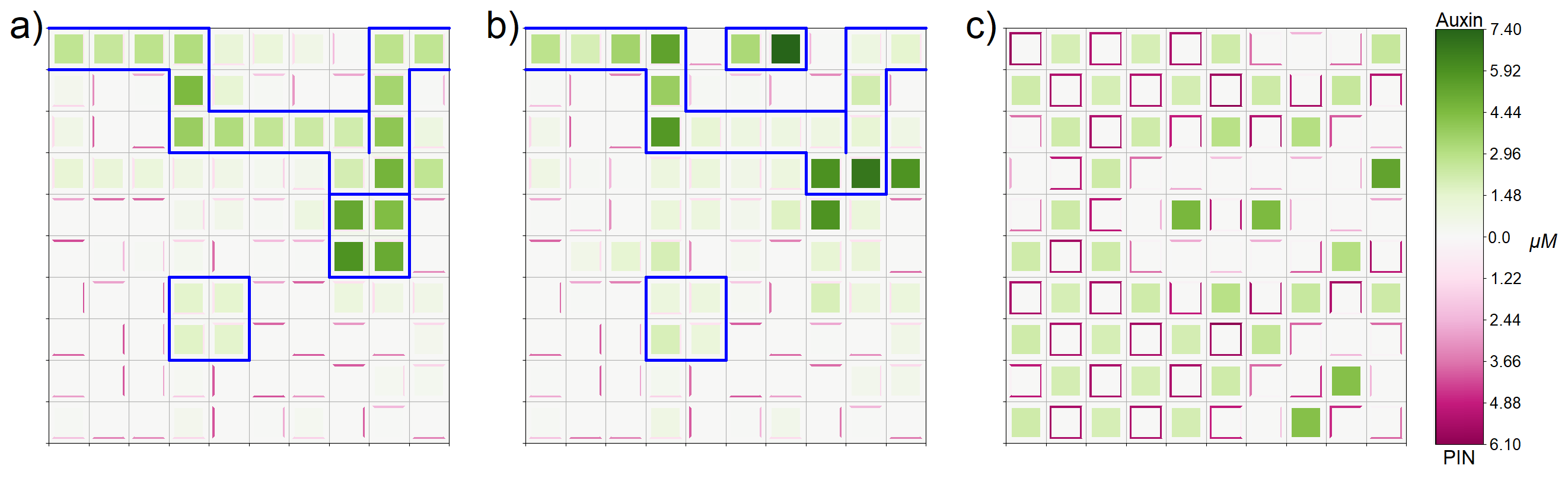}
 \caption{For the symplast-apoplast model the mechanisms of auxin transport and PIN localisation determine the steady-state pattern in auxin distribution. \textbf{a)} for model \eqref{eq:aux_flux-signal_cell},\eqref{eq:Ap-AUX-cell}-\eqref{eq:Ap-Fluxes}b),d) considering non-saturating auxin flux and flux-induced PIN localisation passage and spot patterns emerge similar to model \eqref{eq:aux_flux-signal_cell}-\eqref{eq:aux_flux-signal_fluxes}.  
  \textbf{b)} for model \eqref{eq:aux_flux-signal_cell},\eqref{eq:Ap-AUX-cell}-\eqref{eq:Ap-Fluxes}a),d) considering saturating auxin flux and flux-induced PIN localisation similar patterns to the case with non-saturating auxin flux and flux-induced PIN localisation. \textbf{c)} for model \eqref{eq:aux_flux-signal_cell},\eqref{eq:Ap-AUX-cell}-\eqref{eq:Ap-Fluxes}a),c) considering saturating auxin flux and concentration-induced PIN localisation a pattern of single-cell spots with high auxin concentration  emerges. For model \eqref{eq:aux_flux-signal_cell},\eqref{eq:Ap-AUX-cell}-\eqref{eq:Ap-Fluxes}b),c) considering non-saturating auxin flux and concentration-induced PIN localisation the steady-state distribution is homogeneous (not shown). All parameters are described in Tables~\ref{tab:signal} and \ref{tab:apoplast}, with periodic boundary conditions.}
 \label{fig:Apo3indF-kappa}
\end{figure}

To analyse the effects of the mechanisms for auxin transport through apoplast and PIN localisation on the formation of auxin distribution patterns in a plant tissue we solved model \eqref{eq:aux_flux-signal_cell},  \eqref{eq:Ap-AUX-cell}-\eqref{eq:Ap-Fluxes} numerically on a regular lattice of cells.
When considering non-saturating auxin flux and flux-induced PIN localisation, i.e.~mechanisms of the same form as in model without apoplast, \eqref{eq:aux_flux-signal_cell},\eqref{eq:Ap-AUX-cell}-\eqref{eq:Ap-Fluxes_b},\eqref{eq:Ap-Fluxes_d}, then behaviour is similar to the case without apoplast, with both passage and spot patterns able to emerge, Fig.~\ref{fig:Apo3indF-kappa}a).
When considering non-saturating auxin flux and concentration-induced PIN localisation, \eqref{eq:aux_flux-signal_cell},\eqref{eq:Ap-AUX-cell}-\eqref{eq:Ap-Fluxes_b},\eqref{eq:Ap-Fluxes_c} then the steady state auxin distribution is homogeneous (not shown).
When considering saturating auxin flux and flux-induced PIN localisation, \eqref{eq:aux_flux-signal_cell},\eqref{eq:Ap-AUX-cell}-\eqref{eq:Ap-Fluxes_a},\eqref{eq:Ap-Fluxes_d} then similar patterns emerge as in the case with non-saturating auxin flux and flux-induced PIN localisation, Fig.~\ref{fig:Apo3indF-kappa}b).
When considering saturating auxin flux and concentration-induced PIN localisation, i.e. mechanisms of the same form as considered in \cite{Heisler_M_2006}, \eqref{eq:aux_flux-signal_cell},\eqref{eq:Ap-AUX-cell}-\eqref{eq:Ap-Fluxes_a},\eqref{eq:Ap-Fluxes_c}, then single-cell spots in auxin distribution emerge, Fig.~\ref{fig:Apo3indF-kappa}c).
When varying $\kappa_{p}$, the proportion of auxin-induced PIN localisation in \eqref{eq:Ap-Fluxes_c}, between 0 and 1 and numerically solving \eqref{eq:aux_flux-signal_cell},\eqref{eq:Ap-AUX-cell}-\eqref{eq:Ap-Fluxes_a},\eqref{eq:Ap-Fluxes_c}  we found that heterogeneous spot patterns were only generated for values of $0.5\leq\kappa_{p}\leq 1$, and homogeneous distributions were generated for $0\leq\kappa_{p}<0.5$ (data not shown).
For the simulations in Fig.~\ref{fig:Apo3indF-kappa} we used parameter values in the signalling pathway consistent with those used to generate spot patterns in model \eqref{eq:aux_flux-signal_cell}-\eqref{eq:aux_flux-signal_fluxes}, since when using a smaller value of $\beta_{p}$ consistent with that used to generate passage patterns in model \eqref{eq:aux_flux-signal_cell}-\eqref{eq:aux_flux-signal_fluxes} the higher concentrations of PIN on cell membranes led to pooling of auxin in the apoplast compartments adjacent to high PIN-expressing membranes (data not shown).
  We performed numerical simulations for a set of values of $0 < \beta_{p}\leq 100$, and found that auxin pooling in apoplast compartments occurred for $0 < \beta_{p} \lesssim 50$, whereas auxin concentrations were more realistic for $50\lesssim \beta_{p} \leq 100$. 
Disruption of heterogeneous auxin distribution, and reduced uptake and accumulation of auxin is observed for the symplast-apoplast model for auxin transport \eqref{eq:aux_flux-signal_cell}, \eqref{eq:Ap-AUX-cell}-\eqref{eq:Ap-Fluxes} when AUX1 is not included, $\alpha_{u} = 0$, or PIN is overexpressed compared to AUX1, $\alpha_{p} > \alpha_{u} = 1$,  (data not shown), agreeing with experimental observations \cite{Okada_K_1991,Yang_Y_2006}.

%==============================================================================================
\section{Discussion}
%==============================================================================================

Until recently, one of the main criticisms of the canalisation hypothesis (auxin-flux related localisation of PIN to cell membrane) was its inability to produce spot patterns in auxin distribution in a plant  tissue without any additional assumptions on cell types (e.g.~source/sink cells) \cite{Stoma_S_2008}, despite its accurate capturing of passage patterns~\cite{Feller_C_2015}. Recent results  \cite{Cieslak_M_2015,Hayakawa_Y_2015} have shown  that it is possible to obtain both spot and passage patterns  in auxin distribution  considering the canalisation hypothesis, provided an extra mechanism of either auxin-mediated PIN degradation or auxin self-induced production is considered.  This indicates the importance of intracellular processes for auxin transport in a plant tissue, which in the models  in  \cite{Cieslak_M_2015,Hayakawa_Y_2015} were defined phenomenologically, without considering main biological mechanisms underlying those processes. Many of the mathematical models, e.g.\  in \cite{Hayakawa_Y_2015}, and some results presented here, are restricted to  periodic  boundary conditions, which do not always provide best description of biological systems, however are useful for the analysis of auxin flux in homogeneous domains with no sources or sinks.

In the studies presented here we considered a detailed description of the auxin-related signalling pathway and its influence on PIN dynamics, with a key assumption of the regulation of PIN biosynthesis by ARF and PIN degradation by TIR1. This allowed us to identify that the rate of auxin-signalling-dependent PIN degradation, here represented by binding of PIN and auxin-TIR1, is key to determining the patterns of auxin distribution in plant tissues, Fig.~\ref{fig:Summary2}.

For our model we assumed that the mechanism of auxin-dependent degradation of PIN is similar to the degradation of Aux/IAAs  via the auxin-TIR1 signalling pathway, despite limited biological evidence \cite{Abas_L_2006}. Although we believe that including auxin-dependent PIN degradation is important for the results obtained by our model, it might be possible that this specific mechanism is not essential; for example auxin-dependent PIN degradation has previously been modelled and shown to have a primary role in determining eventual auxin distribution pattern without the need to consider the full auxin signalling pathway \cite{Hayakawa_Y_2015}. Our model does not seem to be able to generate single cell spots when considering flux-induced PIN localisation and this is a key qualitative difference between our model and the model presented  in \cite{Hayakawa_Y_2015}, which is able to generate patterns composed of single cell spots.
Bifurcation analysis for the auxin transport and signalling pathway model presented in this paper suggests that it is possible to obtain spots and passages patterns  for the same parameter values,  with the likelihood of the emergence of passage patterns decreases and the likelihood of the emergence of spot patterns increases as sensitivity of PIN degradation to auxin increases.  This differs from results obtained in   \cite{Hayakawa_Y_2015} where a clear transition from passage-generating to spot-generating parameter regimes as sensitivity of PIN degradation to auxin increases is shown. It  may be possible that adopting a similar mechanism of PIN degradation as in \cite{Hayakawa_Y_2015} in the  model presented here would result in more similar bifurcation behaviour.

Our results on interactions between signalling and transport processes showed that the oscillatory dynamics in auxin concentration are obtained only when considering modified parameter values in the model equations for signalling pathway in all cells in the simulated tissue.
  This suggests that experimentally observed oscillations in auxin responsiveness is due to an oscillatory Aux/IAA negative feedback loop \cite{Middleton_A_2010} and that both the oscillatory feedback loop and PIN-mediated auxin transport through tissue are necessary for the formation of auxin distributions with local oscillations, Fig.~\ref{fig:aux-osc}.
In cells other than those spots which have oscillatory dynamics there are either damped or no oscillations, which is likely due to the very low concentrations of membrane-bound PIN in the membranes of oscillatory cells bordering non-oscillatory cells, i.e.~the oscillatory dynamics of spot cells exert negligible influence on the dynamics of their non-spot neighbours.
It would be of interest to investigate the dynamics of  solutions of  model  \eqref{eq:aux_flux-signal_cell}-\eqref{eq:aux_flux-signal_fluxes} in the oscillatory parameter regime when solved on a realistic plant root geometry.
%Mathematical modelling alongside experimental work has been undertaken to examine oscillations in auxin-dependent gene expression and lateral root formation \cite{Xuan_W_2016}.

It has been observed that mechanical strain of a plasma membrane enhances PIN localisation to the corresponding  membrane \cite{Homann_U_1998}. One model to consider such contributions was proposed in \cite{Hernandez-Hernandez_V_2018}, which modelled PIN localisation on the single cell level using a discrete boolean model, approximating  continuous dynamical system, and predicted that mechanical forces could dominate molecular factors during PIN polarisation.  Our numerical simulation results for the coupled auxin flux  and tissue growth model \eqref{eq:aux_flux-signal_cell}-\eqref{eq:growth_a} indicate
that mechanical forces could dominate the molecular activity since PIN is preferentially polarised, leading to the formation of auxin gradients along the axis of growth, Figs.~\ref{fig:Sig_horizontal},~\ref{fig:Sig_vertical}, whereas the strain-induced localisation of cytoplasmic PIN to the membrane, for membrane strain above a certain threshold, had a qualitative effect on the dynamics of auxin and PIN in a growing plant tissue. Numerical simulation results for mathematical model \eqref{eq:aux_flux-signal_cell}-\eqref{eq:growth_a}, \eqref{eq:strain} also showed that balanced contribution of chemical activities and mechanical forces to the PIN dynamics does not affect the type of patterns in the auxin distribution in a growing tissue. 

Auxin is transported from shoot to root through the stele to the root tip where it is reorganised and then transported back up towards the shoot in the outer cell layers \cite{Grieneisen_V_2007}. This directed auxin flux is commonly known as `reverse-fountain', and has been observed to be essential for root development \cite{Doerner_P_2008}, for example in specifying the quiescent center \cite{Sabatini_S_1999} and root responses to gravitropism \cite{Swarup_R_2005}. Previous mathematical models described reverse flow in auxin patterns by prescribing polarisation of membrane bound PIN \cite{Band_L_2014,Mironova_V_2010,Stoma_S_2008}. Our new mathematical model for auxin transport in a plant tissue, that includes the dynamics of PIN coupled to the auxin-related signalling pathway, auxin flux and tissue growth, is able to generate reverse flow patterns in the auxin distribution from an initial condition that does not have pre-established PIN polarity, Fig.~\ref{fig:Growth2}. Our results suggest a plausible mechanism for the emergence of the `reverse fountain' auxin pattern observed at the root tip: the establishment of the PIN polarity that generates this characteristic auxin distribution is mechanically generated due to the dilution of PIN along growing membranes since when dilution is outweighed by strain-induced localisation the reverse fountain patterns do not emerge. This suggests that in growing tissues strain-induced PIN localisation must be carefully balanced against other mechanisms of PIN localisation to ensure that the correct auxin distributions are established. This hypothesis opens an exciting avenue for further experimental and theoretical investigations of relations between reverse auxin flow and growth processes in plant tissues, especially in plant roots. 

For a model considering auxin flux through the apoplast \eqref{eq:aux_flux-signal_cell},~\eqref{eq:Ap-AUX-cell}-\eqref{eq:Ap-Fluxes} we compared different mechanisms of transmembrane auxin flux and PIN localisation to examine their influence on the formation of auxin patterns. We found for flux-based PIN localisation both passage and spot patterns were able to be produced, however for concentration-based PIN localisation only spots were able to emerge when combined with saturating auxin flux.
When PIN was overexpressed compared to AUX1  homogeneous auxin disruptions were observed, agreeing with experimental results of reduced auxin accumulation in cells and pooling in the apoplast \cite{Okada_K_1991,Yang_Y_2006}.   
Together this suggests some balance between the expressions of PIN and AUX1 is important to facilitate heterogeneous auxin distributions required for stable plant growth. 
These results also suggest that auxin transport through the apoplast has an effect on the dynamics and distribution of auxin and PIN in plant tissues. 
Further experimental and theoretical studies of relations between auxin transport through plasmodesmata and through apoplast are important for a better understanding of auxin dynamics and distribution in plant tissues.

\begin{figure}\centering
 \includegraphics[width=\linewidth]{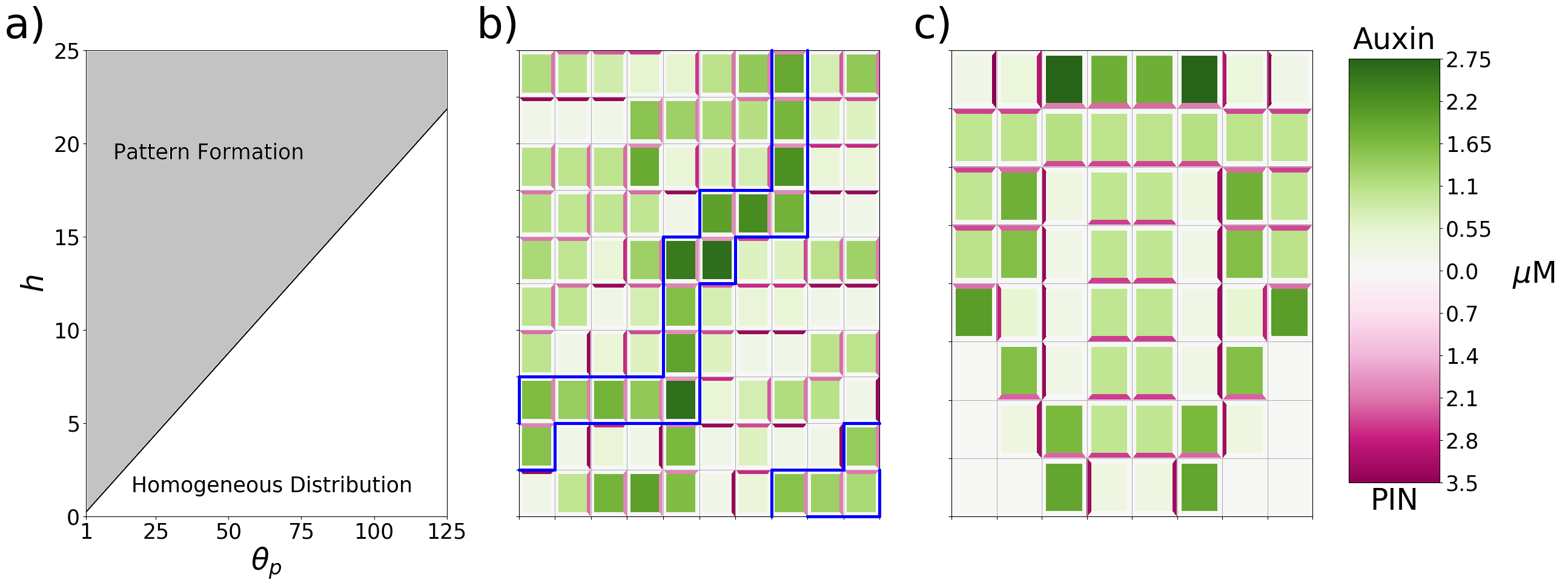}
 \caption{Reduced model \eqref{eq:reduced} has similar dynamics to the full model \eqref{eq:aux_flux-signal_cell}-\eqref{eq:aux_flux-signal_fluxes} for the same parameter values. New parameters in \eqref{eq:reduced} are calculated from previous parameters as detailed. \textbf{a)} zones of pattern formation and homogeneous distribution for parameters $h$ and $\theta_{p}$ resemble those in Fig.~\ref{fig:Summary2}a). Note $\theta_{p} = \beta_{p}/2$. \textbf{b)} Model \eqref{eq:reduced} forms passage patterns with similar characteristics as in Fig.~\ref{fig:Basic_Sig2-Periodic-Spatial-7014}a) with the same parameter values. \textbf{c)} For the same parameter values as in Fig.~\ref{fig:Growth2}c), model \eqref{eq:reduced} generates similar reverse flow patterns at the root tip, where auxin is transported down the root from the central source cells and is then redirect to the outer layers of cells where it is transported back up the root, however the exact alignment of PIN proteins on membranes is different.}
 \label{fig:Reduced}
\end{figure}

We recognise that our model has many components and is more complicated than many other mathematical models which describe the emergence of auxin patterns e.g.~\cite{Feller_C_2015,Feugier_F_2005,Hayakawa_Y_2015}. We chose to include a good level of biological detail in our model so that the dynamics of all components could be predicted and compared with experimental data. However, our model can be simplified significantly by recognising that the dynamics of TIR1-containing components are faster than other reactions and so can be solved for those variables, reducing equations \eqref{eq:aux_flux-signal_cell}, \eqref{eq:aux_flux-signal_membrane} to

\begin{equation}\label{eq:reduced}
 \begin{aligned}
  \frac{dm_{i}}{dt} = \; & \alpha_{m}\dfrac{\phi_{m}f_{i}/\theta_{f} + w_{i}/\theta_{w} + f_{i}^{2}/\psi_{f}}{1 + f_{i}/\theta_{f} + w_{i}/\theta_{w} + g_{i}/\theta_{g} + f_{i}r_{i}/\psi_{g} + f_{i}^{2}/\psi_{f}} - \mu_{m}m_{i},
  \\
  \frac{dr_{i}}{dt} = \; & \alpha_{r}m_{i} - \tilde{\mu}_{r}\frac{\theta_{a}a_{i}\theta_{r}r_{i}}{1 + \theta_{a}a_{i}\left(1 + \theta_{r}r_{i} + \theta_{p}p_{i}\right)} - \beta_{g}r_{i}f_{i} + \gamma_{g}g_{i},
  \\
  \frac{dp_{i}}{dt} = \; & \alpha_{p}m_{i} - \tilde{\mu}_{p}\frac{\theta_{a}a_{i}\theta_{p}p_{i}}{1 + \theta_{a}a_{i}\left(1 + \theta_{r}r_{i} + \theta_{p}p_{i}\right)} - \dfrac{1}{V_{i}}\sum_{i\sim j}S^{m}_{ij}J_{p}^{ij},
  \\
  \frac{df_{i}}{dt} = \; & -2\beta_{f}f_{i}^{2} + 2\gamma_{f}w_{i} - \beta_{g}r_{i}f_{i} + \gamma_{g}g_{i},
  \\
  \frac{dg_{i}}{dt} = \; & \beta_{g}r_{i}f_{i} - \gamma_{g}g_{i},
  \\
  \frac{dw_{i}}{dt} = \; & \beta_{f}f_{i}^{2} - \gamma_{f}w_{i},
  \\
  \frac{da_{i}}{dt} = \; & \alpha_{a} - \mu_{a}a_{i} - \dfrac{1}{V_{i}}\sum_{i\sim j}S^{m}_{ij}J_{a}^{ij},
  \\
  \frac{dP_{ij}}{dt} = \; & J_{p}^{ij},
 \end{aligned}
\end{equation}
where $\tilde{\mu}_{r} = \mu_{r}S_{tot}$, $\tilde{\mu}_{p} = \mu_{p}S_{tot}$, $\theta_{a} = \beta_{a}/\gamma_{a}$, $\theta_{r} = \beta_{r}/(\gamma_{r}+\mu_{r})$, and $\theta_{p} = \beta_{p}/(\gamma_{p}+\mu_{p})$. The reduced model has four fewer variables and four fewer parameters compared to equations~\eqref{eq:aux_flux-signal_cell}, \eqref{eq:aux_flux-signal_membrane}, and combined with equations~\eqref{eq:aux_flux-signal_fluxes} demonstrates similar behaviour as the full model~\eqref{eq:aux_flux-signal_cell}-\eqref{eq:aux_flux-signal_fluxes}   for the same parameter values, Fig.~\ref{fig:Reduced}.
Since it is highly likely that other hormone signalling networks interact with the auxin signalling pathway to maintain auxin distribution patterns \cite{Bishopp_A_2011}, it would be feasible to combine this simplified model with mathematical models of other signalling pathways. For example mathematical modelling of interactions between auxin and cytokinin in plant roots has been investigated in \cite{Mellor_N_2017,Muraro_D_2013a} and it would be of interest to examine the dynamics of an extension of the models presented in this work to include cytokinin.

The influence of auxin on root growth is  highly complex and still not fully understood \cite{Sengupta_D_2018}. The heterogeneous distribution of auxin and its flux through plant  tissues are responsible for the development of tissues including root hair \cite{Zhang_DJ_2018}, vasculature \cite{Marhava_P_2018}, and lateral roots \cite{Li_Z_2018}, shoot branching \cite{Ongaro_V_2008}, flowering \cite{Cheng_Y_2006}, and of course primary root growth \cite{Pelagio-Flores_R_2016,Wakeel_A_2018,Wan_J_2018b}. Auxin influences these developmental processes through the Aux/IAA signalling pathway modulating transcription of relevant proteins, however recent results are also exposing control through non-transcriptional effects downstream of the signalling pathway \cite{Fendrych_M_2018}. Auxin transport depends on the dynamics of PIN polarity \cite{Abas_L_2006}, whereas dynamics of PIN depend on auxin-related cellular signalling processes \cite{Vieten_A_2005}. Considering this nonlinear coupling between signalling processes, auxin transport and PIN dynamics,  we hope that our mathematical model and analysis of nonlinear interactions between auxin flux, cellular signalling pathway, PIN dynamics, and growth, as well as hypotheses resulting from our numerical simulation results  will contribute to a better  understanding of the role of auxin in root development. 
To our knowledge the results presented in this paper on oscillatory auxin transport, comparisons between flux-induced and strain induced PIN localisation and the formation of the reverse fountain without prescribed PIN polarisation patterns are novel.
Our mathematical model for interactions between signalling processes and auxin transport can also be generalised to address  possible direct effect of auxin-related signalling processes on the polarisation of PIN \cite{Sauer_M_2006},  once  more information about  these direct interactions has been found. Further research will also include generalisation of our symplast-apoplast model to include auxin transport through plasmodesmata  and to analyse the effect of auxin on transport through plasmodesmata of various signalling molecules  \cite{Han_X_2014}.

%==============================================================================================

%\begin{acknowledgements}
%H.R.~Allen gratefully acknowledges the support of an EPSRC DTA PhD studentship.
%\end{acknowledgements}

%==============================================================================================
%\bibliographystyle{spmpsci}
\bibliography{References}
%==============================================================================================

\begin{table}\centering
 \begin{tabular}{cc}
  \toprule
   Variable & Meaning\\
  \midrule
   $m_{i}$ & cellular mRNA concentration \\
   $r_{i}$ & cellular Aux/IAA concentration \\
   $s_{i}$ & cellular TIR1 concentration \\
   $c_{i}$ & cellular auxin-TIR1 concentration \\
   $v_{i}$ & cellular Aux/IAA-auxin-TIR1 concentration \\
   $e_{i}$ & cellular PIN-auxin-TIR1 concentration \\
   $f_{i}$ & cellular ARF concentration \\
   $g_{i}$ & cellular ARF-Aux/IAA concentration \\
   $w_{i}$ & cellular ARF\textsubscript{2} concentration \\
   $a_{i}$ & cellular auxin concentration \\
   $p_{i}$ & cellular PIN concentration \\
   $u_{i}$ & cellular AUX1 concentration \\
   $A_{ij}$ & apoplastic auxin concentration \\
   $P_{ij}$ & membrane-bound PIN concentration \\
   $U_{ij}$ & membrane-bound AUX1 concentration \\
  \bottomrule
 \end{tabular}
 \caption{Listing of variable in equationss \eqref{eq:aux_flux-signal_cell}-\eqref{eq:Ap-Fluxes}.}
 \label{tab:variables}
\end{table}

\begin{table}\centering
 \begin{tabular}{clccccc}
  \toprule
   Parameter & Description & Figs.~\ref{fig:Summary2}a),b) & Figs.~\ref{fig:Basic_Sig2-Periodic-Spatial-7014}a), & Figs.~\ref{fig:Basic_Sig2-Periodic-Spatial-7014}b),  & Figs.~\ref{fig:aux-osc},~\ref{fig:Middleton_Growth} \\
   ~ & ~ & ~ & \ref{fig:Sig_horizontal}a),~\ref{fig:Sig_vertical}a), & \ref{fig:Sig_horizontal}b),~\ref{fig:Sig_vertical}b), \\
   ~ & ~ & ~ & \ref{fig:Sig_mech}a),~\ref{fig:Growth2},~\ref{fig:Mech} & \ref{fig:Sig_mech}b),\ref{fig:Apo3indF-kappa} \\
   ~ & ~ & ~ & (Default) & ~ & ~ \\
  \midrule
   $\alpha_{m}$ & maximum mRNA transcription rate, $\mu$M min\textsuperscript{-1} & 0.5,~0-75 & 0.5 & 0.5 & 10 \\
   $\phi_{m}$ & ratio of ARF-dependent to ARF\textsubscript{2}- and double & 0.1 & 0.1 & 0.1 & 0.1 \\
   ~ & ARF-dependent mRNA transcription rates & & \\
   $\theta_{f}$ & ARF-DNA binding threshold, $\mu$M & 1 & 1 & 1 & 1 \\
   $\theta_{w}$ & ARF\textsubscript{2} binding threshold, $\mu$M & 10 & 10 & 10 & 10 \\
   $\psi_{f}$ & double ARF-DNA binding threshold, $\mu$M\textsuperscript{2} & 0.1 & 0.1 & 0.1 & 0.1 \\
   $\theta_{g}$ & ARF + Aux/IAA-DNA binding threshold, $\mu$M & 1 & 1 & 1 & 1 \\
   $\psi_{g}$ & ARF-Aux/IAA-DNA binding threshold, $\mu$M\textsuperscript{2} & 0.1 & 0.1 & 0.1 & 0.1 \\
   $\mu_{m}$ & mRNA decay rate, min\textsuperscript{-1} & 0.05 & 0.05 & 0.05 & 0.05 \\
   $\alpha_{r}$ & Aux/IAA translation rate, min\textsuperscript{-1} & 5 & 5 & 5 & 5 \\
   $\beta_{r}$ & Aux/IAA-auxin-TIR1 binding rate, $\mu$M\textsuperscript{-1} min\textsuperscript{-1} & 5 & 5 & 5 & 5 \\
   $\gamma_{r}$ & Aux/IAA-auxin-TIR1 dissociation rate, min\textsuperscript{-1} & 5 & 5 & 5 & 5 \\
   $\mu_{r}$ & Aux/IAA decay rate, min\textsuperscript{-1} & 5 & 5 & 5 & 5 \\
   $\alpha_{p}$ & PIN translation rate, min\textsuperscript{-1} & 5 & 5 & 5 & 5 \\
   $\beta_{p}$ & PIN-auxin-TIR1 binding rate, $\mu$M\textsuperscript{-1} min\textsuperscript{-1} & 1-250 & 5 & 100 & 5 \\
   $\gamma_{p}$ & PIN-auxin-TIR1 dissociation rate, min\textsuperscript{-1} & 5 & 5 & 5 & 5 \\
   $\mu_{p}$ & PIN decay rate, min\textsuperscript{-1} & 5 & 5 & 5 & 5 \\
   $\beta_{g}$ & ARF-Aux/IAA binding rate, $\mu$M\textsuperscript{-1} min\textsuperscript{-1} & 0.5 & 0.5 & 0.5 & 0.5 \\
   $\gamma_{g}$ & ARF-Aux/IAA dissociation rate, min\textsuperscript{-1} & 5 & 5 & 5 & 5 \\
   $\beta_{a}$ & auxin-TIR1 binding rate, $\mu$M\textsuperscript{-1} min\textsuperscript{-1} & 0.5 & 0.5 & 0.5 & 0.5 \\
   $\gamma_{a}$ & auxin-TIR1 dissociation rate, min\textsuperscript{-1} & 5 & 5 & 5 & 5 \\
   $\beta_{f}$ & ARF dimerisation rate, $\mu$M\textsuperscript{-1} min\textsuperscript{-1} & 0.5,~0.005 & 0.5 & 0.5 & 0.005 \\
   $\gamma_{f}$ & ARF\textsubscript{2} splitting rate, min\textsuperscript{-1} & 5,~0.05 & 5 & 5 & 0.05 \\
   $\alpha_{a}$ & auxin biosynthesis rate, $\mu$M min\textsuperscript{-1} & 0.5 & 0.5 & 0.5 & 0.5 \\
   $\mu_{a}$ & auxin degradation rate, min\textsuperscript{-1} & 0.5 & 0.5 & 0.5 & 0.5 \\
   $S_{tot}$ & total TIR1 present in cell, $\mu$M & 10 & 10 & 10 & 10 \\
   $F_{tot}$ & total ARF present in cell, $\mu$M & 10 & 10 & 10 & 10 \\
  \bottomrule
 \end{tabular}
 \caption{Parameter values for the model of auxin-related signalling pathway \eqref{eq:aux_flux-signal_cell},  calculated from non-dimensionalised values in \cite{Middleton_A_2010}.}
 \label{tab:signal}
\end{table}

\pagebreak

\begin{table}[!ht]\centering
 \begin{tabular}{clcccc}
  \toprule
   Parameter & Description & Fig~\ref{fig:Summary2}a) & Figs.~\ref{fig:Summary2}b), & Figs.~\ref{fig:Basic_Sig2-Periodic-Spatial-7014}, \\
   ~ & ~ & ~ & ~\ref{fig:aux-osc}a),~\ref{fig:Middleton_Growth}b) & ~\ref{fig:aux-osc}b) \\
   ~ & ~ & ~ & ~ & (Default) \\
  \midrule
   $\phi_{A}$ & PIN-dependent auxin transport rate, $\mu$M\textsuperscript{-1} min\textsuperscript{-1} & 0.005 & 0.005 & 0.005 \\
   $\lambda$ & PIN membrane localisation rate, $\mu$m min\textsuperscript{-1} & 0.5 & 0.05 & 0.5 \\
   $\nu$ & strain-dependent PIN localisation rate, $\mu$m min\textsuperscript{-1} & 0 & 0 & 0 \\
   $\delta_{p}$ & PIN membrane dissociation rate, min\textsuperscript{-1} & 0.05 & 0.05 & 0.05 \\
   $h$ & sensitivity of PIN localisation to auxin flux & 0-25 & 50 & 50 \\
   $\theta$ & auxin flux threshold & 2 & 2 & 2 \\
   $\chi$ & maximal cell growth rate, $\mu$m min\textsuperscript{-1} & 0 & 0 & 0 \\
   $\theta_{x}$ & auxin threshold for half-maximal growth rate, $\mu$M & 0 & 0 & 0 \\
  \midrule
  & & Fig.~\ref{fig:Middleton_Growth}a) & Figs.~\ref{fig:Sig_horizontal},~\ref{fig:Sig_vertical}, & Figs.~\ref{fig:Growth2}, \\
  & & & ~\ref{fig:Sig_mech}a),b) & ~\ref{fig:Mech}a),b) \\
  \cline{3-5}
  & & 0.005 & 0.005 & 0.005 \\
  & & 0.05 & 0.5,~0.5,~0.25,~0.1 & 0.5,~0.25,~0.1 \\
  & & 0 & 0,~0,~0.25,~0.4 & 0,~0.25,~0.4 \\
  & & 0.05 & 0.05 & 0.05 \\
  & & 50 & 50 & 50 \\
  & & 2 & 2 & 2 \\
  & & 1 & 1 & 1 \\
  & & 0.8 & 0.8 & 5 \\
  \cline{3-5}
 \end{tabular}
 \caption{Parameter values for flux-based processes in model equations \eqref{eq:aux_flux-signal_membrane}-\eqref{eq:strain}.}
 \label{tab:flux}
\end{table}

\pagebreak

\begin{table}[!ht]\centering
 \begin{tabular}{ccc}
  \toprule
   Parameter & Description & Fig.~\ref{fig:Apo3indF-kappa} \\
  \midrule
   $\alpha_{u}$ & AUX1 biosynthesis rate, $\mu$M min\textsuperscript{-1} & 5 \\
   $\mu_{u}$ & AUX1 degradation rate, min\textsuperscript{-1} & 5 \\
   $\theta_{u}$ & saturation of auxin-induced AUX1 biosynthesis, $\mu$M & 1 \\
   $\phi_{a}$ & auxin membrane permeability, $\mu$m  min\textsuperscript{-1} & 0.55 \\
   $\kappa_{a}^{ef}$ & fraction of protonated auxin in cell & 0.004\\
   $\kappa_{a}^{in}$ & fraction of protonated auxin in wall & 0.24\\
   $\phi_{p}$ & saturating PIN-induced auxin membrane permeability, $\mu$m  min\textsuperscript{-1} & 0.27 \\
   $\tilde{\phi}_{p}$ & PIN-induced auxin membrane permeability, $\mu$m $\mu$M\textsuperscript{-1}  min\textsuperscript{-1} & 0.27 \\
   $\kappa_{p}^{ef}$ & effective PIN-induced auxin efflux & 4.67\\
   $\kappa_{p}^{in}$ & effective PIN-induced auxin influx & 0.034\\
   $\theta_{a}^{p}$ & saturation of PIN-induced auxin transport, $\mu$M & 1 \\
   $\phi_{u}$ & saturating AUX1-induced auxin membrane permeability, $\mu$m min\textsuperscript{-1} & 0.55 \\
   $\tilde{\phi}_{u}$ & AUX1-induced auxin membrane permeability, $\mu$m $\mu$M\textsuperscript{-1} min\textsuperscript{-1} & 0.55 \\
   $\kappa_{u}^{ef}$ & effective AUX1-induced auxin efflux & 0.045\\
   $\kappa_{u}^{in}$ & effective AUX1-induced auxin influx & 3.56\\
   $\theta_{a}^{u}$ & saturation of AUX1-induced auxin transport, $\mu$M & 1 \\
   $\phi_{A}$ & rate of auxin diffusion in apoplast, $\mu$m min\textsuperscript{-1} & 67 \\
   $\omega_{u}$ & rate of AUX1 localisation to membrane, $\mu$m min\textsuperscript{-1} & 0.5 \\
   $\delta_{u}$ & rate of AUX1 dissociation from membrane, min\textsuperscript{-1} & 0.05 \\
   $\omega_{p}$ & maximum rate of PIN localisation to membrane, $\mu$m min\textsuperscript{-1} & 0.5 \\
   $\kappa_{p}$ & fraction of PIN localisation due to auxin feedback & 1 \\
   $\theta_{p}^{a}$ & threshold for half-maximal auxin-dependent PIN localisation, $\mu$M & 1 \\
   $\delta_{p}$ & rate of PIN dissociation from membrane, min\textsuperscript{-1} & 0.05 \\
  \bottomrule
 \end{tabular}
\caption{Parameter values for the auxin flux and PIN  and AUX1 dynamics  considered in model equations, \eqref{eq:Ap-AUX-cell}-\eqref{eq:Ap-Fluxes}, parameters based upon those in \cite{Heisler_M_2006}.}
\label{tab:apoplast}
\end{table}

%==============================================================================================
\end{document}